\documentclass[aps,prb,twocolumn,showpacs,floatfix,amsmath,amssymb]{revtex4}

\usepackage{amsmath,bm,amsfonts}
\usepackage[dvipdfm]{graphicx}

\newfont{\bg}{cmr10 scaled\magstep2}
\newcommand{\bigzerol}{\smash{\hbox{\bg 0}}}
\newcommand{\bigzerou}{\smash{\hbox{\bg 0}}}

\begin{document}

\title{Efficient implementation of the nonequilibrium Green
       function method for electronic transport calculations}

\author{Taisuke Ozaki}
\affiliation{
     Research Center for Integrated Science (RCIS), Japan 
     Advanced Institute of Science and Technology (JAIST), 
     1-1 Asahidai, Nomi, Ishikawa 923-1292 Japan}
\author{Kengo Nishio}
\affiliation{
     Research Institute for Computational Sciences (RICS),
     National Institute of Advanced
     Industrial Science and
     Technology (AIST),
     1-1-1 Umezono, Tsukuba,
     Ibaraki 305-8568, Japan} 
\author{Hiori Kino}
\affiliation{
     National Institute for Material Science (NIMS),
     1-2-1 Sengen, Tsukuba, Ibaraki 305-0047, Japan
}

\date{\today}

\begin{abstract} 

An efficient implementation of the nonequilibrium Green function (NEGF) 
method combined with the density functional theory (DFT) using localized 
pseudo-atomic orbitals (PAOs) is presented for electronic transport 
calculations of a system connected with two leads under a finite bias voltage. 
In the implementation, accurate and efficient methods are developed
especially for evaluation of the density matrix and treatment of 
boundaries between the scattering region and the leads. 
Equilibrium and nonequilibrium contributions in the density matrix 
are evaluated with 
very high precision by a contour integration with a continued fraction 
representation of the Fermi-Dirac function and by a simple quadrature
on the real axis with a small imaginary part, respectively. 
The Hartree potential is computed efficiently  
by a combination of the two dimensional fast Fourier transform (FFT) 
and a finite difference method, and the charge density near the
boundaries is constructed with a careful treatment to avoid the spurious 
scattering at the boundaries. The efficiency of the implementation
is demonstrated by rapid convergence properties of the density matrix.
In addition, as an illustration, our method is applied for zigzag graphene 
nanoribbons, a Fe/MgO/Fe tunneling junction, 
and a LaMnO$_3/$SrMnO$_3$ superlattice, demonstrating its applicability 
to a wide variety of systems.

\end{abstract}


\maketitle

\section{INTRODUCTION}

The non-equilibrium Green function (NEGF) method\cite{Schwinger,Keldysh,Caroli,Datta,Datta2,Jauho}
potentially has several advantages to investigate electronic transport properties 
of nano-scale materials such as single molecules,\cite{Derosa,Kondo} 
atomic wires,\cite{Wei,Grigoriev} 
carbon based materials,\cite{Li,Kim} and thin layers.\cite{Chen,Bulusu} 
The potential advantages are summarized by the following 
features of the NEGF method:
(i) the source and drain contacts are treated based on the same 
theoretical framework as for the scattering region.
\cite{Caroli,Datta,Jauho}
(ii) the electronic structure of the scattering region under 
a finite source-drain bias voltage
is self-consistently determined by combining with first principle electronic
structure calculation methods such as the density functional theory (DFT) 
and the Hartree-Fock (HF) method.
\cite{Taylor,Brandbyge,Novaes,Kim-NEGF,RLi,Rocha,Zhao,Havu,Shima2,Shima3,Wang}
(iii) many body effects in the transport 
properties, e.g., electron-phonon\cite{Asai,Sergueev1,Paulsson,Sergueev2,Shima4,Shima5}
and electron-electron interactions,\cite{Ferretti,Darancet,Thygesen1,Thygesen2} 
might be included through self-energies without largely deviating
the theoretical framework.
(iv) its applicability to large-scale systems can be anticipated, since 
the NEGF method relies practically on the locality of basis functions 
in real space, resulting in computations for sparse matrices.\cite{Nardelli} 
Due to those potential advantages, recently several groups 
have implemented the NEGF method coupled with the DFT or HF method
using atomic-type or the other local basis functions with successful 
applications for calculations of the electronic transport properties.
\cite{Nardelli,Damle,Taylor,Brandbyge,Novaes,Kim-NEGF,RLi,Rocha,Zhao,Havu,Shima2,Shima3,Wang}

However, a highly accurate and efficient implementation method must 
be still developed from the following two reasons:
The first obvious reason is to extend the applicability of the NEGF method
to large-scale systems. The efficient implementation might lead to 
more challenging applications of the NEGF method to very large-scale 
complicated systems.
The majority part in the computational effort of the NEGF method mainly 
comes from the evaluation of the density matrix which is decomposed into 
the evaluation of Green functions and numerical integrations. 
Thus, the efficient calculation of the part is a key factor 
for extending the applicability to large-scale systems. 
Nevertheless, accurate and efficient methods for evaluating 
density matrix within the NEGF method have not been fully developed, 
although several methods have been already proposed.
\cite{Taylor,Brandbyge,Williams} 
To extend the applicability of the NEGF method to large-scale systems, 
a remarkably efficient method that we have recently developed\cite{Ozaki-FD}
will be applied for the problem in this study, and we will show that 
the new method is much faster than the other method.\cite{Brandbyge} 
The second reason is that spurious scattering should be negligible
when the NEGF method is extended to include the many body effects beyond 
the one particle picture.\cite{Asai,Sergueev1,Paulsson,Sergueev2,Shima4,Shima5,Ferretti,Darancet,Thygesen1,Thygesen2} 
The spurious scattering accompanied by the inaccurate implementation
might make the many body effects indistinct in the electronic transport 
properties. One can imagine that the spurious scattering can be easily produced
in the NEGF method, since unlike the conventional band structure calculations
NEGF has to be evaluated by a patch work that the self consistent field (SCF) 
calculations of the source and drain leads are performed beforehand, 
and the calculated results are incorporated in the NEGF calculations
through the self energy and the boundary conditions between the scattering
region and leads. Therefore, a careful treatment to handle the boundary 
conditions should be developed to avoid the spurious scattering. 

In this paper, to address the above two issues we present an accurate
and efficient implementation of the NEGF method, in combination with DFT using 
pseudo-atomic orbitals (PAOs) and pseudopotentials, using a contour integration
method which is based on a continued 
fraction representation of the Fermi-Dirac function. 
For the accurate treatment of the boundary conditions between the 
scattering region and leads, we also develop a method for 
calculating the Hartree potential by a combination 
of the two dimensional fast Fourier transform (FFT) and a finite difference 
method so that the boundary condition can be correctly reproduced. 
In addition, we discuss a careful treatment to construct the charge density 
near the boundaries. The efficiency and accuracy of our implementation 
are demonstrated by several numerical test calculations on convergence 
of the density matrix.

This paper is organized as follows: 
In Sec.~II the details of our implementation for treating the equilibrium 
state of the scattering region are discussed by focusing on the evaluation
of the equilibrium density matrix, the treatment for constructing the charge 
density near the boundaries, and an efficient method for calculating
the Hartree potential. 
In Sec.~III our implementation for the nonequilibrium state is described.
In Sec.~IV we demonstrate the accuracy and efficiency of the 
implementation by a series of numerical calculations and 
several applications.
In Sec.~V our implementation of the NEGF method is summarized.

\section{EQUILIBRIUM STATE}

Since most of practical aspects in the implementation of the NEGF
method coupled with DFT using the 
localized PAOs\cite{Ozaki-PAO1,Ozaki-PAO2,Ozaki-PAO3} 
appear in the ground state calculation of the system 
at equilibrium, we start our discussion from the electronic 
structure calculation of the equilibrium ground state 
by using the Green function method with DFT. 

\subsection{Equilibrium Green function (EGF)}

Let us consider a system where one dimensional infinite cells
are arranged with two dimensional periodicity as shown in Fig.~1.
Throughout the paper, we assume that the electronic transport
along the {\bf a}-axis is of interest, and that the two dimensional 
periodicity spreads over the {\bf bc}-plane.
The one dimensional infinite cell consists of the central region
denoted by $C_0$ and the cells denoted by $L_i$ and $R_i$, 
where $i=0,1,2\cdots$. All the cells $L_i$ and $R_i$, 
arranged semi-infinitely, contain the same number of atoms
with the same structural configuration, respectively, 
but the cells $L_i$ and $R_i$ can be different from each other. 
In the equilibrium state with a common chemical potential
everywhere in the system, the electronic structure of the 
system may be determined by DFT.\cite{Hohenberg,Kohn}
\begin{figure}[t]
    \centering
    \includegraphics[width=8.0cm]{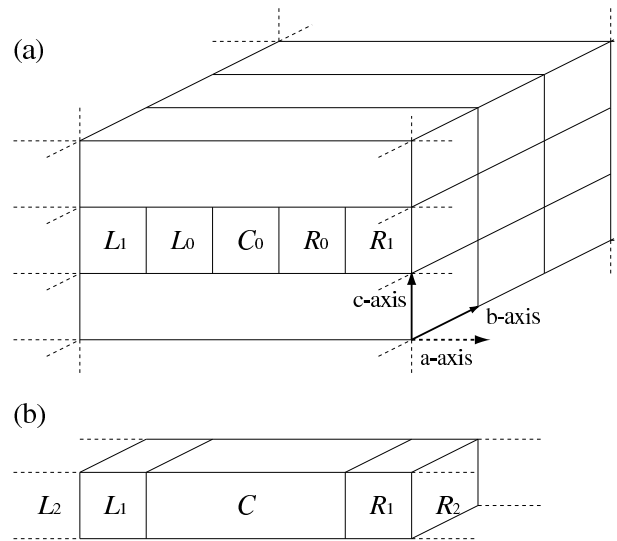}
    \caption{(a) Configuration of the system, treated by the NEGF method, 
             with infinite left and right leads
             along the {\bf a}-axis under a two dimensional periodic 
             boundary condition on the {\bf bc}-plane. 
             (b) One dimensional system compacted from the configuration 
             of (a) by considering 
             the periodicity on the {\bf bc}-plane, where the region $C$
             is an extended central region consisting 
             of $C_{0}$, $L_{0}$, and $R_{0}$. 
             }
\end{figure}
Due to the periodicity of the {\bf bc}-plane the one-particle Kohn-Sham (KS)
wave function in the system is expressed by the Bloch function 
on the {\bf bc}-plane using PAOs $\phi_{i\alpha}$ located on 
site $\tau_i$ as:
\begin{eqnarray}
  \psi_{\sigma\nu}^{(\bf k)}({\bf r})
   =   
  \frac{1}{\sqrt{N}}
  \sum_{\rm n}
  {\rm e}^{i{\bf k}\cdot {\bf R}_{\rm n}}
  \sum_{i\alpha}c_{\sigma\nu,i\alpha}^{(\bf k)}
  \phi_{i\alpha}({\bf r}-\tau_{i}-{\bf R}_{\rm n}),  
  \label{eq:c1-1}
\end{eqnarray}
where $\sigma$, $\nu$, $i$, and $\alpha$ are indices for the spin, 
eigenstate, site, and basis orbital, respectively.
The lattice vector ${\bf R}_{\rm n}$ and the Bloch wave vector ${\bf k}$ 
are given by ${\bf R}_{\rm n} = l_b{\bf b}+l_c{\bf c}$, where ${\bf b}$
and ${\bf c}$ are the lattice vectors, and 
${\bf k}= k_b{\bf \tilde{b}}+k_c{\bf \tilde{c}}$,
where ${\bf \tilde{b}}$ and ${\bf \tilde{c}}$ are the 
reciprocal lattice vectors, respectively. 
The summation over $i$ and $\alpha$ is considered for all the basis orbitals
in the one dimensional infinite cell, which indicates no periodicity 
along the ${\bf a}$-axis. Considering the variation 
of the total energy, within the conventional DFT, 
of the system expressed by the KS wave function Eq.~(\ref{eq:c1-1})
with respect to coefficients $c$ we obtain the following KS matrix equation: 
\begin{eqnarray}
  H_{\sigma}^{(\bf k)} c_{\sigma\nu}^{(\bf k)}
  = \varepsilon_{\sigma\nu}^{(\bf k)} 
    S^{(\bf k)} 
    c_{\sigma\nu}^{(\bf k)},
  \label{eq:c1-2}
\end{eqnarray}
where $c_{\sigma\nu}^{(\bf k)}$ is a column vector consisting 
of the coefficients $\{ c_{\sigma\nu,i\alpha}^{(\bf k)} \}$.
The Hamiltonian $H_{\sigma}^{(\bf k)}$ and overlap matrices 
$S^{(\bf k)}$ are given by 
\begin{eqnarray}
  H_{\sigma,i\alpha j\beta}^{(\bf k)}
 &=& \sum_{\rm n} {\rm e}^{i{\bf k}\cdot {\bf R}_{\rm n}}
     h_{\sigma,i\alpha j\beta,{\bf R}_{\rm n}},
  \label{eq:c1-3}\\
  S_{i\alpha j\beta}^{(\bf k)}
 &=& \sum_{\rm n} {\rm e}^{i{\bf k}\cdot {\bf R}_{\rm n}}
     s_{i\alpha j\beta,{\bf R}_{\rm n}},
  \label{eq:c1-4}
\end{eqnarray} 
where $h_{\sigma,i\alpha j\beta,{\bf R}_{\rm n}}$ and 
$s_{i\alpha j\beta,{\bf R}_{\rm n}}$ are the Hamiltonian 
and overlap matrix elements between two basis functions
$\phi_{i\alpha}({\bf r}-\tau_{i})$ 
and 
$\phi_{j\beta}({\bf r}-\tau_{j}-{\bf R}_{\rm n})$,
respectively. 
The overlap matrix arises from the non-orthogonality of the PAO basis 
functions.\cite{Ozaki-PAO1,Ozaki-PAO2,Ozaki-PAO3} 
Now we consider an extended central region $C$ composed 
of the regions $C_0$, $L_0$, and $R_0$ as shown in Figs.~1 (a) and (b). 
The extension of the central region $C_0$ is made 
so that the relaxation of electronic structure around the interfaces 
between the leads, $L_0$ and $R_0$, and the central region $C_0$ can 
be allowed. In addition, we impose two conditions:
\renewcommand{\labelenumi}{(\roman{enumi})}
\begin{enumerate}
\item 
The localized basis orbitals
$\phi$ in the region $C_0$ overlap with those in the regions $L_0$ and $R_0$,
but do not overlap with those in the regions $L_1$ and $R_1$. 

\item 
The localized basis orbitals $\phi$ in the $L_i$ ($R_i$) region 
has no overlap with basis orbitals in the cells beyond the nearest 
neighboring cells $L_{i-1}$ ($R_{i-1}$) and $L_{i+1}$ ($R_{i+1}$).

\end{enumerate}
In our implementation the basis functions are strictly localized in 
real space because of the generation of basis orbitals by a confinement
scheme.\cite{Ozaki-PAO1,Ozaki-PAO2,Ozaki-PAO3}
 Therefore, once the localized basis orbitals with specific
cutoff radii are chosen for each region, the two conditions 
can be always satisfied by just adjusting the size of the unit cells
for $L_i$ and $R_i$. This is a benefit in the use of the strictly 
localized basis orbitals compared to other local basis orbitals 
such as Slater- and Gaussian-type orbitals. 
In the use of the strictly localized 
basis orbitals, the eigenvalue problem in the Hilbert space spanned 
by the basis orbitals are solved without introducing any cutoff 
scheme. On the other hand, in the Green function method
the matrix elements of the Hamiltonian and overlap matrices 
have to be truncated so as to satisfy the 
above two conditions in case of the other localized basis orbitals 
with the small but long tail. With the above two conditions (i) and (ii), 
the Hamiltonian matrix given by Eq.~(\ref{eq:c1-3}) is written 
by a block tridiagonal form as follows:
\begin{eqnarray}
  H_{\sigma}^{(\bf k)}
  = 
  \left(
   \begin{array}{ccccc}
 \ddots & \ddots & & & \bigzerou  \\
 \ddots & H_{\sigma,L_1}^{(\bf k)} & H_{\sigma,L_1 C}^{(\bf k)} &&   \\
        & H_{\sigma,CL_1}^{(\bf k)}& H_{\sigma,C}^{(\bf k)} & H_{\sigma,CR_1}^{(\bf k)} & \\
        &  & H_{\sigma,R_1C}^{(\bf k)} & H_{\sigma,R_1}^{(\bf k)} & \ddots \\
    \bigzerol & & & \ddots & \ddots \\
   \end{array}
  \right),
  \label{eq:c1-5}
\end{eqnarray}
where $H_{\sigma,C}^{(\bf k)}$, $H_{\sigma,L_1}^{(\bf k)}$, and 
$H_{\sigma,R_1}^{(\bf k)}$ are Hamiltonian matrices of the central $C$,
left $L_{1}$ and right $R_{1}$ regions of which matrix size are 
the same as the number of basis orbitals $N_{C}$, $N_{L}$, and
$N_{R}$ in the regions $C$, $L_1$, and $R_1$, respectively.
The other block components in Eq.~(\ref{eq:c1-5}) are the Hamiltonian matrices
connecting two regions among the regions, and these matrix sizes are deduced
from those of the two regions. Also the completely same structure is 
found in the overlap matrix. 
Thus, the electronic structure of the system given by Fig.~1(a) can be obtained
by solving the one dimensional block chain model, being ${\bf k}$-dependent, 
given by Fig.~1(b) and the corresponding Eq.~(\ref{eq:c1-2}). 
By noting
$G_{\sigma}^{(\bf k)}(Z)(ZS^{(\bf k)}-H_{\sigma}^{(\bf k)})={\rm I}$ 
and making use of the block 
tridiagonal form of the Hamiltonian and overlap matrices, 
the Green function of the central region $C$ can be written by
\begin{eqnarray}
  G_{\sigma,C}^{(\bf k)}(Z) = 
  \left(ZS_{C}^{(\bf k)}-H_{\sigma,C}^{(\bf k)}
       -\Sigma_{\sigma,L}^{(\bf k)}(Z)
       -\Sigma_{\sigma,R}^{(\bf k)}(Z)\right)^{-1}
  \label{eq:c1-6}
\end{eqnarray}
with self energies $\Sigma_{\sigma,L}^{(\bf k)}(Z)$ and 
$\Sigma_{\sigma,R}^{(\bf k)}(Z)$ defined by 
\begin{eqnarray}
 \nonumber
 \Sigma_{\sigma,L}^{(\bf k)}(Z)
 &=& (ZS_{CL_1}^{(\bf k)}-H_{\sigma,CL_1}^{(\bf k)})
   \times\\
  &&
     G_{\sigma,L}^{(\bf k)}(Z)
    (ZS_{L_1C}^{(\bf k)}-H_{\sigma,L_1C}^{(\bf k)}),
  \label{eq:c1-7}\\
 \nonumber
 \Sigma_{\sigma,R}^{(\bf k)}(Z)
 &=& (ZS_{CR_1}^{(\bf k)}-H_{\sigma,CR_1}^{(\bf k)})
   \times\\
  &&
     G_{\sigma,R}^{(\bf k)}(Z)
    (ZS_{R_1C}^{(\bf k)}-H_{\sigma,R_1C}^{(\bf k)}),
  \label{eq:c1-8}
\end{eqnarray}
where $G_{\sigma,L}^{(\bf k)}(Z)$ and $G_{\sigma,R}^{(\bf k)}(Z)$
are surface Green functions of the left and right regions. 

It is worth pointing out that there is a total energy functional 
which can be variationally minimized with respect to charge density $n$
if Eq.~(\ref{eq:c1-6}) is self-consistently solved. 
The details of derivation for the functional is given in Appendix A. 

\subsection{Surface Green function}

In general, the surface Green function 
$G_{\sigma,{\rm s}}^{(\bf k)}(Z)$ is defined by 
$G_{\sigma,{\rm s}}^{(\bf k)}(Z)\equiv (ZS_{\rm s}-H_{\rm s})^{-1}$, 
where $S_{\rm s}$ and $H_{\rm s}$ are the Hamiltonian and overlap matrices 
for the lead regions, and the suffix, ${\rm s}$, is $L$ or $R$.
It is noted that due to the two conditions (i) and (ii) mentioned above 
the Hamiltonian and overlap matrices for the lead 
regions can be written by a block tridiagonal form as follows:
\begin{eqnarray}
  H_{\rm s}
  &=& 
  \left(
   \begin{array}{cccc}
     H_{11} & H_{12} & & \bigzerou \\
     H_{21} & H_{22} & H_{23} & \\
     & H_{32} & H_{33} & \ddots \\
    \bigzerol &  & \ddots & \ddots \\
   \end{array}
  \right),
  \label{eq:c1-9}\\
  S_{\rm s}
  &=& 
  \left(
   \begin{array}{cccc}
     S_{11} & S_{12} & & \bigzerou \\
     S_{21} & S_{22} & S_{23} & \\
     & S_{32} & S_{33} & \ddots \\
    \bigzerol &  & \ddots & \ddots \\
   \end{array}
  \right),
  \label{eq:c1-10}
\end{eqnarray}
where the Hamiltonian can be spin and ${\bf k}$-dependent,
while the indices for them are omitted for simplification 
of the notation, and also the index $i$ appearing in $H_{ij}$
corresponds to the cell number for the lead $L_{i}$ or $R_{i}$.
It seems to be difficult to directly diagonalize 
the KS equation Eq.~(\ref{eq:c1-2})
for the one dimensional block chain model because of the infinite dimension
of the matrices. However, instead by focusing on only the central region
one can evaluate the Green function of the central region as a rather small
problem of $N_C \times N_C$ in size. The effect of semi infinite regions $L$
and $R$ are included through the corresponding self energies 
$\Sigma_{\sigma,L}^{(\bf k)}(Z)$ and $\Sigma_{\sigma,R}^{(\bf k)}(Z)$.
In order to practically calculate the Green function of the central region
given by Eq.~(\ref{eq:c1-6}), we introduce an approximation
where the regions $L_{i} (i=1,2,\cdots)$ are all equivalent to each other 
with respect to the spatial charge distribution, the KS Hamiltonian,
and the relevant density matrix which are calculated in advance
by adopting the system of which unit cell is $L_1$ and by using
the conventional band structure calculation. 
The same approximation also applies for the regions $R_{i} (i=1,2,\cdots)$. 
Strictly speaking, the assumption is not correct, since the charge 
distribution must be affected by the interaction between 
the central region $C$ and the regions $L_i$ ($R_i$). 
However, if the size of the unit vector along the ${\bf a}$-axis
for the regions $L_0$ and $R_0$ in the extended central region $C$ 
is large enough, the assumption will be asymptotically correct 
as the unit vector becomes larger.  
The approximation enables us to evaluate the surface Green function 
by the iterative method.\cite{Lopez} 
The efficient iterative scheme can be performed by the following procedure:
\begin{eqnarray}
  a_{i} &=& \epsilon_{i}^{-1}\alpha_{i},
  \label{eq:c1-11}\\  
  b_{i} &=& \epsilon_i^{-1}\beta_{i},
  \label{eq:c1-12}\\
  \epsilon_{{\rm s},i+1} &=& \epsilon_{{\rm s},i} - \alpha_{i}b_{i},
  \label{eq:c1-13}\\
  \epsilon_{i+1} &=& \epsilon_{i}-\beta_{i}a_{i}-\alpha_{i}b_{i},
  \label{eq:c1-14}\\
  \alpha_{i+1} &=& \alpha_{i}a_{i},
  \label{eq:c1-15}\\
  \beta_{i+1} &=& \beta_{i}b_{i}
  \label{eq:c1-16}
\end{eqnarray}
with a set of initial values 
\begin{eqnarray}
  \epsilon_{{\rm s},0} &=& ZS_{11} - H_{11},
  \label{eq:c1-17}\\  
  \epsilon_{0}   &=& ZS_{11} - H_{11},
  \label{eq:c1-18}\\  
  \alpha_{0} &=& -(ZS_{12} - H_{12}),
  \label{eq:c1-19}\\  
  \beta_{0}  &=& -(ZS_{21} - H_{21}).
  \label{eq:c1-20}
\end{eqnarray}
By the iterative calculation, in most cases 
the inverse of $\epsilon_{{\rm s},i}$ 
rapidly converges at a part of the surface Green function:
\begin{eqnarray}
  G_{{\rm s},11} = \lim_{i\to \infty} \epsilon_{{\rm s},i}^{-1},
  \label{eq:c1-21}
\end{eqnarray}
where $G_{{\rm s},11}$ is the (1,1) block element of the surface
Green function $(ZS_{\rm s}-H_{\rm s})^{-1}$. The (1,1) block element
$G_{{\rm s},11}$ of $N_L \times N_L$ (or $N_R \times N_R$) in size
has all the necessary information to calculate the 
self-energies $\Sigma_{\sigma,L}^{(\bf k)}(Z)$ and 
$\Sigma_{\sigma,R}^{(\bf k)}(Z)$, since
there is no contribution from the other block elements
because of the two conditions (i) and (ii) mentioned above.
In practice, the convergence in the iterative calculation is 
very fast and the Frobenius norm, defined by
$
  (\sum_{l,l'}\vert 
  (\epsilon_{{\rm s},i+1})_{ll'}
 -(\epsilon_{{\rm s},i})_{ll'}\vert^2)^{1/2}
$,
of $10^{-5}$ (eV) is obtained by typically only 7 iterations.

\subsection{Equilibrium density matrix}

One of practical difficulties in the implementation of
the Green function method is how the equilibrium density 
matrix is evaluated efficiently and accurately.\cite{Taylor,Brandbyge,Williams,Nicholson,Goedecker,Wildberger,Drchal,Matsubara} 
In our implementation, the equilibrium density matrix is 
highly efficiently computed using the contour integration method
with a special treatment of the Fermi-Dirac function $f$.\cite{Ozaki-FD} 
If the Hamiltonian and overlap matrices associated with Eq.~(\ref{eq:c1-6}) 
are ${\bf k}$-dependent, it turns out that the spectrum function
in the Lehmann representation of the central Green function 
is complex number in general as discussed in Appendix B 
of Ref.~\cite{Ozaki-FD}.
Then, the density matrix $\rho_{\sigma,{\bf R}_{\rm n}}^{\rm (eq)}$, 
where one of the associated basis orbitals is in the central cell 
and the other is in the cell denoted by ${\bf R}_{\rm n}$, 
is given by making use of
both the retarded and advanced Green functions 
$G_{\sigma,C}^{(\bf k)}(E+i0^+)$ 
and 
$G_{\sigma,C}^{(\bf k)}(E-i0^+)$
as 
\begin{eqnarray}
   \rho_{\sigma,{\bf R}_{\rm n}}^{\rm (eq)}
    = 
    \frac{1}{V_{\rm c}}
    \int_{\rm BZ} dk^3
    \left(
    \rho_{\sigma,+}^{(\bf k)} - \rho_{\sigma,-}^{(\bf k)}
    \right)
    {\rm e}^{-i{\bf k}\cdot {\bf R}_{\rm n}}
  \label{eq:c1-22}
\end{eqnarray}
with 
\begin{eqnarray}
    \rho_{\sigma,\pm}^{(\bf k)}
    = 
    \frac{i}{2\pi}
    \int_{-\infty}^{\infty}
    dE
    G_{\sigma,C}^{(\bf k)}(E\pm i0^+)
    f(E-\mu),
  \label{eq:c1-23}
\end{eqnarray}
where $V_{\rm c}$ is the volume of the unit cell, $\int_{\rm BZ}$ 
represents the integration over the first Brillouin zone,
$0^+$ a positive infinitesimal, and $\mu$ a chemical potential. 
The integration over ${\bf k}$-space is 
numerically performed by using the Monkhorst-Pack mesh.\cite{MK} 
It is also noted that the phase factor 
${\rm e}^{-i{\bf k}\cdot {\bf R}_{\rm n}}$ appears through 
Eq.~(\ref{eq:c1-1}). 
If the Hamiltonian and overlap matrices are 
${\bf k}$-independent, Eq.~(\ref{eq:c1-22}) can be simplified 
into a well-known formula:
\begin{eqnarray}
   \rho_{\sigma,0}^{\rm (eq)}
   = 
  {\rm Im}\left[
    -\frac{1}{\pi}
    \int_{-\infty}^{\infty}
    dE
    G_{\sigma,C}(E+i0^+)
    f(E-\mu)
    \right].
  \label{eq:c1-24}
\end{eqnarray}
In case of the ${\bf k}$-independent problem, the simplified formula is used,
since the number of the evaluation of the Green function is reduced 
by half. 

For the efficient integration in Eq.~(\ref{eq:c1-23}), in our implementation
the Fermi-Dirac function is expressed by a continued fraction 
representation derived from a hypergeometric function\cite{Ozaki-FD}
so that the structure of poles can be suitable for the integration 
associated with the Green function as follows:
\begin{eqnarray}
  \nonumber
  \frac{1}{1+\exp(x)}
  &=&
    \frac{1}{2} -  
    \frac{\strut \frac{x}{4}}
      {\displaystyle 1 + \frac{\strut (\frac{x}{2})^2}
      {\displaystyle 3 + \frac{\strut (\frac{x}{2})^2}
      {\displaystyle 5 + \frac{\strut (\frac{x}{2})^2}
      {\displaystyle \frac{\strut \cdots}
      {(2M-1)+}_{\ddots}
      }}}}\\
   &=&
   \frac{1}{2} + 
   \sum_{p=1}^{\infty}\frac{R_{p}}{x-iz_p} 
   +
   \sum_{p=1}^{\infty}\frac{R_{p}}{x+iz_p},
  \label{eq:c1-25}
\end{eqnarray}
where $x=\beta(z-\mu)$ with $\beta=\frac{1}{k_{\rm B}T}$, 
$T$ is electronic temperature, $z$ and $x$ are complex variables.
Also, $z_p$ and $R_p$ are the poles and the associated residues of the 
continued fraction representation Eq.~(\ref{eq:c1-25}), which are 
obtained via an eigenvalue problem derived from Eq.~(\ref{eq:c1-25}).\cite{Ozaki-FD} 
Since all the $z_p$ are real numbers, the poles $iz_p$ are 
located on the imaginary axis. 
One may find an interesting distribution of the poles on the complex plane 
that the interval between neighboring poles are uniformly located 
up to about 61~\% of the total number of poles on the half complex plane 
with the same interval $2\pi$, 
and from then onward it increases very rapidly as the distance between the pole
and the real axis increases. The structure of the poles in Eq.~(\ref{eq:c1-25})
allows us to efficiently evaluate Eq.~(\ref{eq:c1-23}) because of 
the asymptotic change, $1/Z$, of the Green function
in the faraway region of the real axis. In other words, 
the denser poles are allocated for the rapidly varying range of 
the Green function, and the coarser for smoothly. 
In addition, there is no ambiguity in the choice of the path 
in the contour integration unlike the other schemes,\cite{Taylor,Brandbyge,Williams}
which is one of advantages in our method.
By terminating the summation of Eq.~(\ref{eq:c1-25}) 
at a finite number of poles $N_p$, 
the integration of Eq.~(\ref{eq:c1-23}) can be performed by the contour 
integration where the poles in the upper and lower half planes are taken 
into account for the terms with plus and minus signs in Eq.~(\ref{eq:c1-23}),
respectively, and the explicit formula is given by 
\begin{eqnarray}
   \rho_{\sigma,\pm}^{(\bf k)}
    &=& 
   \pm\frac{1}{4}\mu^{({\bf k},0)}_{\sigma}
   \mp\frac{1}{\beta} \sum_{p=1}^{N_p} 
      G_{\sigma,C}^{(\bf k)}(\alpha_p)R_p,
  \label{eq:c1-26}
\end{eqnarray}
where $\alpha_{p}=\mu\pm i\frac{z_p}{\beta}$, 
and $\mu^{({\bf k},0)}_{\sigma}$ is the zeroth order 
moment of the Green function $G_{\sigma,C}^{(\bf k)}$.
The zeroth order moment $\mu^{({\bf k},0)}_{\sigma}$
is easily calculated by 
$\mu^{({\bf k},0)}=iR~G_{\sigma,C}^{(\bf k)}(iR)$ 
which can be derived from the moment representation of 
the Green function,\cite{Ozaki-FD}
where $R$ is a large real number and in this study 
$10^{10}$ (eV) is used in order to make the higher order 
moments negligible. 
Although the number of poles in the summation of Eq.~(\ref{eq:c1-26})
required for the sufficient convergence depends on 
the electronic temperature $T$, 
the fully convergent result within double precision is achieved by 
the use of only 100 poles in case of $T=600~K$ as shown later.

If forces on atoms are calculated based on the conventional DFT scheme
using the non-orthogonal basis orbitals,\cite{Brandbyge} 
the evaluation of the energy density matrix $e_{\sigma}$ 
is needed. In Appendix B, we derive the calculation scheme 
of the equilibrium energy density matrix $e_{\sigma}^{\rm (eq)}$ 
based on the contour integration method.

\subsection{Charge density near the boundary}

Even though the basis functions we used are strictly localized 
in real space, there is the nonnegligible contribution for 
the charge density near the boundary between the central and 
lead regions from the basis functions located in the lead regions.
Note that any treatment for the contribution to the charge density 
has not been clarified 
in the other implementations.\cite{Taylor,Brandbyge,Novaes,Kim-NEGF,RLi,Rocha,Zhao}
Thus, we carefully calculate the charge density in the central 
region by considering three contributions:
\begin{eqnarray}
  n_{\sigma}({\bf r}) = 
       n_{\sigma}^{\rm (cc)}({\bf r})
     + 2n_{\sigma}^{\rm (sc)}({\bf r})
     + n_{\sigma}^{\rm (ss)}({\bf r}),
  \label{eq:c1-43}
\end{eqnarray}
where the suffix, s, is $L$ or $R$, and 
$n_{\sigma}^{\rm (cc)}({\bf r})$, 
$n_{\sigma}^{\rm (sc)}({\bf r})$,
and 
$n_{\sigma}^{\rm (ss)}({\bf r})$
are the charge densities contributed from the basis functions
located in the central, the lead and central, and the lead regions, 
respectively. Note that the summation over s is not required 
in Eq.~(\ref{eq:c1-43}) because of the conditions (i) and (ii). 
Each charge contribution is explicitly given by 
\begin{eqnarray}
  \nonumber
  n_{\sigma}^{\rm (cc)}({\bf r}) 
  &=& 
  \sum_{\rm n}
  \sum_{i\alpha,j\beta}
  \rho_{\sigma,i\alpha,j\beta{\bf R}_{\rm n}}^{\rm (eq)}\\
  &&
  \times  
  \phi_{i\alpha}({\bf r}-\tau_{i}) 
  \phi_{j\beta}({\bf r}-\tau_{j}-{\bf R}_{\rm n}),
  \label{eq:c1-44}\\
  n_{\sigma}^{\rm (sc)}({\bf r}) 
  &=& 
  \nonumber
  \sum_{\rm n,n'}
  \sum_{i\alpha,j\beta}
  \rho_{\sigma,i\alpha{\bf R}_{\rm n},j\beta{\bf R}_{\rm n'}}^{\rm (sc)}\\  
  \nonumber
  &&
  \times  
  \phi_{i\alpha}({\bf r}-\tau_{i}-({\bf R}_{\rm n}\pm {\bf a}))\\
  &&
  \times  
  \phi_{j\beta}({\bf r}-\tau_{j}-{\bf R}_{\rm n'}),
  \label{eq:c1-45}\\
  \nonumber
  n_{\sigma}^{\rm (ss)}({\bf r}) 
  &=& 
  \nonumber
  \sum_{\rm n,n'}
  \sum_{i\alpha,j\beta}
  \rho_{\sigma,i\alpha{\bf R}_{\rm n},j\beta{\bf R}_{\rm n'}}^{\rm (ss)}\\  
  \nonumber
  &&
  \times  
  \phi_{i\alpha}({\bf r}-\tau_{i}-({\bf R}_{\rm n}\pm {\bf a}))\\
  &&
  \times  
  \phi_{j\beta}({\bf r}-\tau_{j}-({\bf R}_{\rm n'}\pm {\bf a})),
  \label{eq:c1-46}
\end{eqnarray}
where {\bf a} is the lattice vector of the unit cell for the 
$L_0$ or $R_0$ region along the {\bf a}-axis. 
The displacement of $-{\bf a}$ ($+{\bf a}$) denotes that the basis function 
is placed in the $L_1$ ($R_1$) region in the configuration shown in Fig.~1.
The charge density given by Eq.~(\ref{eq:c1-44}) is calculated 
by the equilibrium density matrix
$\rho_{\sigma,i\alpha,j\beta{\bf R}_{\rm n}}^{\rm (eq)}$
given by Eq.~(\ref{eq:c1-22}). Although Eq.~(\ref{eq:c1-44})
can give a finite electron density on the outside of the central 
cell with ${\bf R}_{\rm n}=0$ because of the overlap of basis 
functions, the contribution is reflected in the central cell 
with ${\bf R}_{\rm n}=0$ by considering the periodicity on
the {\bf bc}-plane.
In the nonequilibrium case, the equilibrium density matrix is 
only replaced by the nonequilibrium density matrix which will be 
discussed in the next section.
Each term in the summations for the two contributions 
$n_{\sigma}^{\rm (sc)}({\bf r})$ and 
$n_{\sigma}^{\rm (ss)}({\bf r})$ survive only if the overlap 
of the associated two basis orbitals is not zero in the central 
region. 
Since the original central region $C_0$ is extended by adding 
the $L_0$ and $R_0$ regions, it is expected that 
the density matrix elements
$\rho_{\sigma,i\alpha{\bf R}_{\rm n},j\beta{\bf R}_{\rm n'}}^{\rm (sc)}$
and 
$\rho_{\sigma,i\alpha{\bf R}_{\rm n},j\beta{\bf R}_{\rm n'}}^{\rm (ss)}$
in Eqs.~(\ref{eq:c1-45}) and ~(\ref{eq:c1-46}) are close to 
those of the leads in the equilibrium condition. 
Therefore, the density matrix elements of the leads calculated 
by the conventional band structure calculations are used for 
Eqs.~(\ref{eq:c1-45}) and (\ref{eq:c1-46}).
Due to the treatment 
the charge densities $n_{\sigma}^{\rm (sc)}({\bf r})$ 
and 
$n_{\sigma}^{\rm (ss)}({\bf r})$ 
are independent of the SCF iteration 
so that for the computational efficiency they can be 
computed on a numerical mesh and stored before the SCF iteration.
The factor 2 for $n_{\sigma}^{\rm (sc)}({\bf r})$ in Eq.~(\ref{eq:c1-43}) 
is due to taking account of the contribution from
$n_{\sigma}^{\rm (cs)}({\bf r})$,
while the factor does not appear for $n_{\sigma}^{\rm (ss)}({\bf r})$, 
since all the paired terms are included by the double summation in 
Eq.~(\ref{eq:c1-46}). 

We add a note that the same consideration has to be applied even for the 
calculation of the density of states (DOS). In this case, 
the contribution from off-diagonal block Green functions connecting 
the central and lead regions should be added to the DOS of the central 
region $C$ calculated by $G_{\sigma,C}^{(\bf k)}(Z)$. 
The off-diagonal block Green functions can be calculated 
from $G_{\sigma,C}^{(\bf k)}(Z)$ and the surface Green functions,
$G_{\sigma,L_1L_1}^{(\bf k)}(Z)$ and 
$G_{\sigma,R_1R_1}^{(\bf k)}(Z)$, by making use of the identity 
$G_{\sigma}^{(\bf k)}(Z)(ZS^{(\bf k)}-H_{\sigma}^{(\bf k)})={\rm I}$
as follows:
\begin{eqnarray}
 G_{\sigma,C{\rm s}_1}^{(\bf k)}(Z)
   =
  -G_{\sigma,C}^{(\bf k)}(Z)
  \left(
    ZS_{C{\rm s}_1}-H_{\sigma,C{\rm s}_1}^{(\bf k)}
  \right)
  G_{\sigma,{\rm s}_1{\rm s}_1}^{(\bf k)}(Z),
 \quad
 \label{eq:c1-63}
\end{eqnarray}
where s is $L$ or $R$.

\subsection{Hartree potential with the boundary condition}

The Hartree potential in the central region is calculated 
under the boundary condition that the Hartree potential 
at the boundary between the central $C$ and $L_1 (R_1)$ 
regions is same as that of the lead,
where the Hartree potential in both the lead regions 
is calculated using the conventional band structure calculation before 
the calculation of the infinite $chain$ in Fig.~1(b) using the 
Green function. In our implementation, 
the Hartree potential for the central 
region with the boundary condition is efficiently 
evaluated by a combination of the two dimensional FFT
and a finite difference method, while other schemes are used 
in the other implementations.\cite{Taylor,Brandbyge}  
The majority part of the Hartree potential in our treatment
is accurately calculated by considering the neutral atom 
potential which is the sum of the local potential of the 
pseudopotential and the Hartree potential by the confined 
charge for the neutralization.\cite{Ozaki-VNA}
The neutral atom potential depends on only the atomic 
structure and atomic species, 
and has no relation with the boundary condition. 
The effect of the relaxation of charge distribution on the 
Hartree potential is taken into account by the remaining 
minority part of the Hartree potential $\Delta V_{\rm H}$
given by 
\begin{eqnarray}
  \nabla^2 \Delta V_{\rm H}({\bf r})
  =  
  -4\pi\Delta n({\bf r}),
  \label{eq:c1-35}
\end{eqnarray}
where $\Delta n({\bf r})$ is defined by the difference between 
the electron density $n({\bf r}) (\equiv \sum_{\sigma}n_{\sigma}({\bf r}))$ 
calculated by the Green function
and the atomic electron density\cite{NA-expln} 
$n^{(a)}({\bf r})$ 
calculated by superposition of each atomic electron density 
$n^{(a)}_i({\bf r})$ at atomic site $i$ as follows: 
\begin{eqnarray}
  \Delta n({\bf r}) = n({\bf r}) - n^{(a)}({\bf r}).
  \label{eq:c1-36}
\end{eqnarray}
The Fourier transformation of Eq.~(\ref{eq:c1-35}) 
on the yz plane, corresponding to the {\bf bc}-plane depicted in 
Fig.~1(a), yields
\begin{eqnarray}
  \left(
  \frac{d^2}{dx^2}
   - 
   {\bf G}^2
  \right)
  \Delta\tilde{V}_{\rm H}(x,{\bf G}) 
  =  
  -4\pi\Delta \tilde{n}(x,{\bf G}) 
  \label{eq:c1-37},
\end{eqnarray}
where ${\bf G}\equiv G_b{\bf \tilde{b}}+G_c{\bf \tilde{c}}$ with 
integer numbers $G_b$ and $G_c$. By approximating the second 
derivative in Eq.~(\ref{eq:c1-37}) with the simplest finite 
difference, 
\begin{eqnarray}
  \nonumber
   \frac{d^2 \Delta \tilde{V}_{\rm H}(x_n)}{dx^2} 
    \simeq 
    \frac{\Delta \tilde{V}_{\rm H}(x_{n+1})
         -2\Delta \tilde{V}_{\rm H}(x_{n})
         +\Delta \tilde{V}_{\rm H}(x_{n-1})}{(\Delta x)^2},\\
  \label{eq:c1-37A}
\end{eqnarray}
we obtain a simultaneous linear equation 
$A \Delta \tilde{V} = B$  of $(N_a-1)\times (N_a-1)$ in size with a tridiagonal 
matrix defined by 
\begin{eqnarray}
  \nonumber
   A_{nn}     &=& 2+(\Delta x)^2 {\bf G}^2,\\
   A_{n(n+1)} &=& A_{(n+1)n} = -1 
  \label{eq:c1-38}
\end{eqnarray}
and a vector defined by 
\begin{eqnarray}
   \nonumber
    B_{1} &=& 
     4\pi (\Delta x)^2 \Delta \tilde{n}(x_1,{\bf G})
             +\Delta \tilde{V}_{\rm H}(x_{0},{\bf G}),\\
   \nonumber
    B_{n} &=& 
     4\pi (\Delta x)^2 \Delta \tilde{n}(x_n,{\bf G}),\\
   \nonumber
    B_{N_a-1} &=& 
     4\pi (\Delta x)^2 \Delta \tilde{n}(x_{N_a-1},{\bf G})
             +\Delta \tilde{V}_{\rm H}(x_{N_a},{\bf G}),\\
  \label{eq:c1-39}
\end{eqnarray}
where $\Delta x$ is the interval between neighboring points $x_n$ and $x_{n+1}$,
$n$ runs from $1$ to $N_a-1$, and $\Delta \tilde{V}_{\rm H}(x_{0},{\bf G})$ and 
$\Delta \tilde{V}_{\rm H}(x_{N_a},{\bf G})$ are the boundary conditions.
Since the lattice vector of the extended central region $C$ along 
the {\bf a}-axis is divided by $N_a$ for the discretization, 
the positions $x_0$ and $x_{N_a}$ are situated at the boundary, 
along the {\bf a}-axis, of the unit cell of the central region $C$. 
Thus, $\Delta \tilde{V}_{\rm H}(x_{0},{\bf G})$ and 
$\Delta \tilde{V}_{\rm H}(x_{N_a},{\bf G})$ 
can be calculated from the Hartree potential at the left boundaries 
along the {\bf a}-axis of the left and right leads, respectively. 
After solving the simultaneous linear equations for all the 
${\bf G}$ points, the inverse two dimensional FFT on 
the {\bf bc}-plane yields the difference Hartree potential 
$\Delta V_{\rm H}$ under the boundary conditions.  
It is apparent that there is no ambiguity for the inclusion 
of the boundary conditions in the method.
Although the treatment can be easily extended to the higher 
order finite difference to the second derivative, we restrict 
ourselves to the simplest case in the implementation because 
of its sufficient accuracy. Employing FFT for two dimensional
Fourier transformation, the whole computational effort to solve 
the Poisson equation Eq.~(\ref{eq:c1-35}) under the boundary 
conditions is estimated to be 
$\sim N_a \times N_b\log(N_b)\times N_c\log(N_c)$
which is slightly superior to that of the three dimensional FFT,
where $N_a$, $N_b$, and $N_c$ are the number of meshes for the
discretization along the ${\bf a}$-, ${\bf b}$-, and ${\bf c}$-axes.

\subsection{Hamiltonian and overlap matrix elements}

Our implementation of the NEGF method with DFT is based on the 
strictly localized PAOs\cite{Ozaki-PAO1,Ozaki-PAO2,Ozaki-PAO3}
and a norm-conserving pseudopotential method.\cite{TM}
Within the scheme, 
the calculation of the matrix elements such as the overlap 
and kinetic energy integrals, consisting of two center integrals, 
is performed using a Fourier transform method,\cite{Sankey}
while the other matrix elements for $V_{\rm xc}$ and $\Delta V_{\rm H}$,
which cannot be decomposed into two center integrals, are evaluated 
by the numerical integration on the regular mesh in real space.\cite{SIESTA}
The further details on how the elements of the Hamiltonian and overlap 
matrices are calculated can be found in Ref.~\cite{Ozaki-VNA}.

In addition to the above evaluation of the Hamiltonian and overlap 
matrix elements, the Hamiltonian matrix elements associated with 
the basis orbitals situated at near the boundary are treated in 
a special way as explained below.
If the tails of two basis orbitals located on atoms
in the central region $C$ go beyond the boundary between the central 
and the lead regions, the associated Hamiltonian matrix element
is replaced by the corresponding element in the lead region
calculated by the conventional band structure calculation.
The case can happen only if the two basis orbitals are located in the 
region $L_0$($R_0$) because of the condition (i) so that the 
replacement of the Hamiltonian matrix element can be always possible.
The replacement is made by assuming that the potential profile near 
the boundary is similar to that near the boundary between the regions
$L_1(R_1)$ and $L_2(R_2)$, and can be justified if the size of the
region $L_0(R_0)$ is large enough.

\subsection{Charge mixing}

Compared to conventional band structure calculations, it seems that 
the NEGF method tends to suffer from difficulty in obtaining the SCF  
convergence. Our observation in several cases suggests that the difficulty 
may come from charge sloshing along the {\bf a}-axis during the SCF iteration. 
The difference Hartree potential $\Delta V_{\rm H}$ change largely 
by imposition of the boundary condition even for a small variation 
in the charge density distribution, resulting in a serious charge sloshing
along the {\bf a}-axis. Thus, we consider suppression of the charge sloshing 
along the {\bf a}-axis by introducing the following weight factor $w$: 
\begin{eqnarray}
  w(x_i,{\bf G}) = g(x_i)\left(
                   \frac{\vert {\bf G}\vert^2 + \kappa_1\vert {\bf G}_0\vert^2}
                        {\vert {\bf G}\vert^2 + \kappa_0\vert {\bf G}_0\vert^2}
                        \right),
  \label{eq:b1-1}
\end{eqnarray}
where $\vert {\bf G}_0\vert$ is a smaller one of either $\vert \tilde{b}\vert$ 
or $\vert\tilde{c}\vert$, and $\kappa_0$ and $\kappa_1$ are adjustable parameters, 
while keeping $\kappa_0<\kappa_1$. The prefactor $g(x_i)$ is given by 
\begin{eqnarray}
   g(x_i)=\left\{ \begin{array}{ll}
                 \xi\left|d_L(x_i) - d_R(x_i)\right| + 1 & {\bf G}=0\\
                  1 & {\rm otherwise}
                 \end{array}
          \right.
  \label{eq:b1-2}
\end{eqnarray}
with definitions:
\begin{eqnarray}
   d_L(x_i) &=& \sum_{k=0}^{i-1}\Delta \tilde{n}_{\rm H}(x_k, {\bf G}=0),\\ 
   d_R(x_i) &=& \sum_{k=i+1}^{N_a-1}\Delta \tilde{n}_{\rm H}(x_k, {\bf G}=0),
  \label{eq:b1-3}
\end{eqnarray}
where $\xi$ is an adjustable parameter. Noting that 
$\Delta \tilde{n}_{\rm H}(x_k, {\bf G}=0)$
is the number of difference electron density of each layer indexed by $k$, and 
that the Coulomb potential induced by each charged layer depends linearly on
the distance from the layer, one can notice that $\left|d_L(x_i) - d_R(x_i)\right|$
is proportional to the electric field at position $i$. 
Therefore, Eq.~(\ref{eq:b1-1}) takes charge density under a large electric 
field into significant account in addition to the suppression of the charge 
sloshing in the {\bf bc}-plane in a sense by the Kerker method.\cite{Kerker} 
In our implementation, 
the weight factor given by Eq.~(\ref{eq:b1-1}) is combined with 
the Kerker method\cite{Kerker} and the residual minimization method in 
a direct inversion iterative subspace (RMM-DIIS)\cite{Kresse} with substantial 
improvement. 

A technical remark should also be added to avoid a local trap problem 
in the SCF calculation. In systems having a long {\bf a}-axis, a unphysical 
charge distribution, corresponding to a large charge separation in real space, 
tends to be obtained even after achieving the self consistency. 
In this case the Hamiltonian of the central region $C$ at the first SCF iteration,
which are calculated via superposition of atomic charge density, 
is far from the self-consistently converged one, while the Hamiltonian matrix
used in the calculation for the self-energy is determined in a self-consistent 
manner beforehand. The inconsistency between the two matrices tends to produce 
a unphysical charge distribution at the first SCF iteration. 
Once the situation happens at the first SCF iteration, in many cases the 
electronic structure keeps trapped during the subsequent SCF iteration, which
is a serious problem in practical applications.
However, the local trap problem can be overcome by a simple scheme that 
the first few SCF iterations are performed by using 
the conventional band scheme and then onward the solver is switched from 
the band scheme to the NEGF method. In the band structure calculation for the 
first few iterations, such a unphysical charge distribution does not appear 
due to no self-energy involved. In most cases we find that the simple scheme 
works well to avoid the local trap problem in the SCF convergence.

\section{NONEQUILIBRIUM STATE}

\subsection{Nonequilibrium density matrix}

Based on the NEGF theory mainly developed 
by Schwinger\cite{Schwinger} and Keldysh\cite{Keldysh}, 
the density matrix in the nonequilibrium state of the central 
region is evaluated by\cite{Taylor,Brandbyge,RLi,Rocha} 
\begin{eqnarray}
  \rho_{\sigma,{\bf R}_{\rm n}}^{\rm (neq)} 
  =   
  \rho_{\sigma,{\bf R}_{\rm n}}^{\rm (eq)}  
  + 
  \Delta\rho_{\sigma,{\bf R}_{\rm n}}.
  \label{eq:c1-40}
\end{eqnarray}
In addition to the equilibrium density matrix 
$\rho_{\sigma,{\bf R}_{\rm n}}^{\rm (eq)}$ given 
by Eq.~(\ref{eq:c1-22}), a correction term defined by 
\begin{eqnarray}
  \Delta\rho_{\sigma,{\bf R}_{\rm n}}
  =   
  \frac{1}{V_{\rm c}}
    \int_{\rm BZ} dk^3
    \Delta\rho_{\sigma}^{(\bf k)}
    {\rm e}^{-i{\bf k}\cdot {\bf R}_{\rm n}}
  \label{eq:c1-41}
\end{eqnarray}
is taken into account, where $\Delta\rho_{\sigma}^{(\bf k)}$
is defined by 
\begin{eqnarray}
  \nonumber
  \Delta\rho_{\sigma}^{(\bf k)}
 &=& 
   \frac{1}{2\pi}
    \int_{-\infty}^{\infty}
    dE
    G_{\sigma,C}^{(\bf k)}(E+i\epsilon) 
    \Gamma_{\sigma,{\rm s}_{1}}^{(\bf k)}(E)\\
   && \qquad\qquad \times
    G_{\sigma,C}^{(\bf k)}(E-i\epsilon)
    \Delta f(E)
  \label{eq:c1-42}
\end{eqnarray}
with 
\begin{eqnarray}
  \Gamma_{\sigma,{\rm s}_{1}}^{(\bf k)}(E)
  =   
   i\left( 
      \Sigma_{\sigma,{\rm s}_{1}}^{(\bf k)}(E+i\epsilon)
      -
      \Sigma_{\sigma,{\rm s}_{1}}^{(\bf k)}(E-i\epsilon)
    \right)
  \label{eq:c1-51}
\end{eqnarray}
and 
\begin{eqnarray}
  \Delta f(E)
  =   
    f(E-\mu_{{\rm s}_1})
    - 
    f(E-\mu_{{\rm s}_2}).
  \label{eq:c1-52}
\end{eqnarray}
Either the left $\mu_L$ or right chemical potential $\mu_R$ 
which is lower than the other is used for the calculation of the equilibrium 
density matrix in Eq.~(\ref{eq:c1-40}). 
As well, in Eqs.~(\ref{eq:c1-42}) and (\ref{eq:c1-52})
the chemical potentials are 
given by the rule that $s_1=R$ and $s_2=L$ if $\mu_{L}<\mu_{R}$, 
and $s_1=L$ and $s_2=R$ if $\mu_{R}\le\mu_{L}$. 
Starting from the NEGF theory, the formula, Eq.~(\ref{eq:c1-40}), 
may be derived by introducing two assumptions. 
The first assumption is that the occupation of the wave functions 
incoming from the left (right) region still obeys the Fermi-Dirac 
function with the left (right) chemical potential even in the 
central region. The assumption can be justified 
within at least the one-particle picture, since the same result
can be obtained from the Lippmann-Schwinger equation 
for a non-interacting system.\cite{Datta2,Taylor,Paulsson2}
The second assumption is that in the central region the states 
in the energy regime below the lower chemical potential is in equilibrium 
due to the other physical obstacles such as the 
electron-phonon\cite{Asai,Sergueev1,Paulsson,Sergueev2,Shima4,Shima5} 
and electron-electron interactions,\cite{Ferretti,Darancet,Thygesen1,Thygesen2}
which are not considered explicitly 
in our implementation, although the definite role of those obstacles 
is obscure as for the occupation of the states.
The second assumption allows electrons to occupy in 
highly localized states below the lower chemical potential
through the first term of Eq.~(\ref{eq:c1-40}).
Only the states in the energy regime between two chemical potentials
are treated as in the nonequilibrium condition in which the contribution 
of the wave functions incoming from the lead with the higher chemical 
potential is taken into account to form the correction term given 
by Eq.~(\ref{eq:c1-41}).

The integrand in Eq.~(\ref{eq:c1-42}) is not analytic apart from 
the real axis, since the integrand is a function of both 
$Z(=E+i\epsilon)$ and $Z^*$. Thus, one cannot apply the contour 
integration method that we use for the equilibrium density matrix. 
Instead, a simple rectangular quadrature scheme is applied to 
the integration of Eq.~(\ref{eq:c1-42}) on the real axis 
with a small imaginary part $\epsilon$. 
Since the integrand contains the difference
between two Fermi-Dirac functions, the energy range for the integration
can be effectively reduced to a narrow range that the difference is larger 
than a threshold, where the threshold of $10^{-12}$ is used in this study.
With the threshold and the step width of 0.01 (eV), the number of meshes 
on the real axis is 152 for $\vert \mu_L - \mu_R \vert=0.1$ (eV)
at $T=300~K$. The convergence speed depends on the shape of the integrand
and how large $\epsilon$ is employed for smearing the integrand,
which will be discussed later. 

\subsection{Source-drain and gate bias voltages}

The source-drain bias voltage applied to the left and right leads is easily incorporated
by adding a constant electric potential $V_{\rm b}$ to the Hartree 
potential in the right lead region. The effect of the bias voltage
appears at three places. The first effect is that the Hamiltonian matrix in the 
right region given by Eq.~(\ref{eq:c1-9}) is replaced using
Eq.~(\ref{eq:c1-10}) as 
\begin{eqnarray}
  H_R \to H_R + V_{\rm b} S_R.
 \label{eq:c1-53}
\end{eqnarray}
This can be easily confirmed by noting that the matrix elements
for the constant potential $V_{\rm b}$ is $V_{\rm b} S_R$. 
The off-diagonal block elements
$H_{\sigma,CL_1}^{(\bf k)}$, $H_{\sigma,L_1C}^{(\bf k)}$
$H_{\sigma,CR_1}^{(\bf k)}$, and $H_{\sigma,R_1C}^{(\bf k)}$ 
appearing Eqs.~(\ref{eq:c1-7}) and (\ref{eq:c1-8}) are also 
replaced in the same way. 
The second effect is that the chemical potential of the right lead is 
replaced as 
\begin{eqnarray}
  \mu_R \to \mu_R + V_{\rm b}.
 \label{eq:c1-54}
\end{eqnarray}
The treatment is made so that the first replacement can be regarded 
as just shifting the origin of energy in the right lead. 
The last effect is that the boundary condition in Eq.~(\ref{eq:c1-39})
is replaced as 
\begin{eqnarray}
 \Delta V_{\rm H}(x_{N_a},{\bf G}) \to \Delta V'_{\rm H}(x_{N_a},{\bf G}),
 \label{eq:c1-55}
\end{eqnarray}
where $\Delta V'_{\rm H}(x_{N_a},{\bf G})$ is calculated by the Fourier
transformation on the {\bf bc}-plane for 
$\Delta V_{\rm H}$ at $x_{N_a}$ plus $V_{\rm b}$.
Since only the difference of the bias voltages applied to the 
left and right leads affects the result, one can consider the 
replacements on only the right lead at the three places as shown above.
It is noted that the replacement by Eq.~(\ref{eq:c1-55}) 
corresponds to adding a linear potential $ax+b$ to 
the Hartree potential in the central region $C$, where $a$ and $b$ 
are determined by the boundary conditions: 
$\Delta V'_{\rm H}(x_{N_a},{\bf G})$ and $\Delta V_{\rm H}(x_{0},{\bf G})$.

In our implementation, the gate voltage $V_{\rm g}(x)$ is treated by adding 
an electric potential defined by 
\begin{eqnarray}
 V_{\rm g}(x) = 
 V_{\rm g}^{(0)} \exp\left[-\left(\frac{x-x_{\rm c}}{d}\right)^{8}\right],
 \label{eq:c1-56}
\end{eqnarray}
where $V_{\rm g}^{(0)}$ is a constant value corresponding to the gate voltage, 
$x_{c}$ the center of the region $C_0$, and $d$ the length of 
the unit vector along {\bf a}-axis for the region $C_0$. Due to the form 
of Eq.~(\ref{eq:c1-56}), the applied gate voltage affects mainly the region 
$C_0$ in the central region $C$.
The electric potential may resemble the potential produced by the image 
charges.\cite{Liang}

\subsection{Transmission and current}

The spin resolved transmission is evaluated by the Landauer formula 
for the non-interacting central region $C$ connected with two leads: 
\begin{eqnarray}
 T_{\sigma}(E) &=& 
 \frac{1}{V_{\rm c}}
 \int_{\rm BZ} dk^3
  T^{(\bf k)}_{\sigma}(E)
 \label{eq:c1-57},
\end{eqnarray}
where $T^{(\bf k)}_{\sigma}(E)$ is the spin and $\bf k$-resolved 
transmission defined by 
\begin{eqnarray}
 T^{(\bf k)}_{\sigma}(E) 
  &=& 
 {\rm Tr}
 \left[
   \Gamma_{\sigma,{\rm L}_{1}}^{(\bf k)}(E)
    G_{\sigma,C}^{(\bf k)}(E+i\epsilon)
   \nonumber
 \right.\\
  && 
  \left.
  \qquad
  \times
  \Gamma_{\sigma,{\rm R}_{1}}^{(\bf k)}(E)
   G_{\sigma,C}^{(\bf k)}(E-i\epsilon) 
  \right] 
 \label{eq:c1-58}.
\end{eqnarray}
Using the transmission formula, the current is evaluated by  
\begin{eqnarray}
 I_{\sigma} = \frac{e}{h}
 \int dE T_{\sigma}(E) \Delta f(E).
 \label{eq:c1-59}
\end{eqnarray}
The formula can be derived by starting from a more general formula of 
the current for the interacting central region $C$ and by replacing 
the involved Green functions with the non-interacting Green functions 
as shown by Meir and Wingreen.\cite{Meir} 
We perform the integration in Eq.~(\ref{eq:c1-59}) on the real 
axis with a small imaginary part $\epsilon$ by the same way as 
for the nonequilibrium density matrix of Eq.~(\ref{eq:c1-42}).

\section{NUMERICAL RESULTS}

\subsection{Computational details}

All the calculations in this study were performed by 
the DFT code, OpenMX.\cite{openmx} 
The PAOs centered on atomic sites are used as basis 
functions.\cite{Ozaki-PAO1,Ozaki-PAO2,Ozaki-PAO3} 
The PAO basis functions we used, generated 
by a confinement scheme,\cite{Ozaki-PAO1,Ozaki-PAO2} are 
specified by H5.5-s2, C4.5-s2p2, O5.0-s2p2d1, Fe5.0-s2p2d1, Mg5.5-s2p2,
La7.0-s3p2d1f1, Sr7.0-s3p2d1f1, and Mn6.0-s3p2d2, 
where the abbreviation of basis functions, 
such as C4.5-s2p2, represents that C stands for the atomic symbol, 
4.5 the cutoff radius (bohr) in the generation by the confinement 
scheme, and s2p2 means the employment of two primitive 
orbital for each of s- and p-orbitals. 
Norm-conserving pseudopotentials are used in 
a separable form with multiple projectors to replace the deep 
core potential into a shallow potential.\cite{TM} Also, a local density
approximation (LDA) to the exchange-correlation potential 
is employed.\cite{LDA}, while a generalized gradient approximation (GGA)\cite{GGA} 
is used only for calculations of the LaMnO$_3/$SrMnO$_3$ superlattice. 
The real space grid techniques are used with 
the cutoff energies of 120-200 Ry in numerical integrations 
and the solution of Poisson equation using FFT.\cite{SIESTA}
In addition, the projector expansion method is employed in the calculation
of three-center integrals for the deep neutral atom potentials.\cite{Ozaki-VNA}

\subsection{Convergence properties}

\begin{figure}[t]
    \centering
    \includegraphics[width=8.0cm]{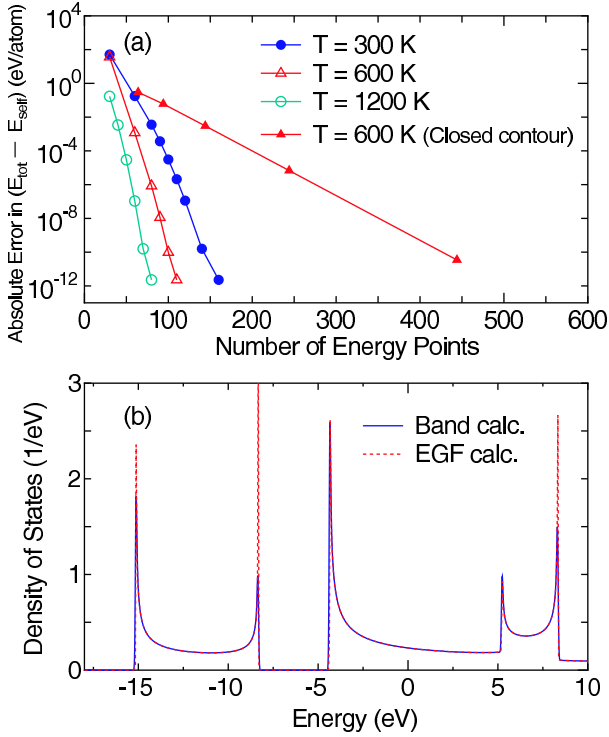}
    \caption{(Color online)(a) Absolute error 
              in ($E_{\rm tot}-E_{\rm self}$) per atom
              of a linear carbon chain with a bond length of 
              1.4~\AA~under the zero bias voltage
              at electronic temperature of 300, 600, and 1200 K, 
              where the regions $L_0$, $R_0$, and $C_0$ contain four
              carbon atoms, respectively. For comparison the same 
              calculation {\it (closed contour)} using a closed contour 
              method\cite{Brandbyge,ClosedContour} is also shown for T=600~K. 
              The definitions of $E_{\rm tot}$ and $E_{\rm self}$
              are found in Appendix A. The reference values are obtained 
              from calculations with a large number of poles.
              (b) Total DOS of the carbon linear chain,
              calculated by the conventional band structure 
              calculation (solid line) and the EGF method (dotted line), 
              under the zero bias voltage at 300 K, 
              where 160 poles are used for the integration of the 
              equilibrium density matrix. 
              It is hard to distinguish two lines due to the 
              nearly equivalent results.}
\end{figure}

The accuracy and efficiency of the implementation are mainly 
determined by the evaluation of density matrix given by Eq.~(\ref{eq:c1-40})
which consists of two contributions: the equilibrium and 
non-equilibrium terms given by Eqs.~(\ref{eq:c1-22}) and (\ref{eq:c1-41}), 
respectively.
In this subsection, we discuss the convergence properties of the 
equilibrium and non-equilibrium terms in the density matrix as 
a function of numerical parameters.

\begin{figure}[t]
    \centering
    \includegraphics[width=8.0cm]{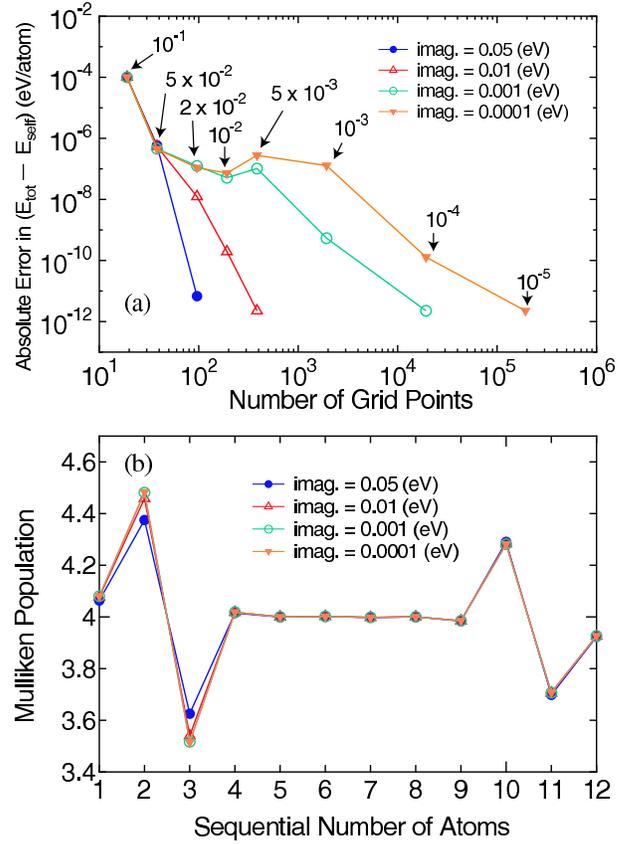}
    \caption{
  (Color online) 
  (a) Absolute error in ($E_{\rm tot}-E_{\rm self}$) per
  atom of the same linear carbon chain as in Fig.~2, 
  calculated by various imaginary parts, under a finite bias 
  voltage of 0.5 V, where the other calculation conditions are 
  same as for Fig.~2(b).
  The number pointed by 
  the arrow denotes a grid spacing (eV) corresponding to the number 
  of grid points. 
  The reference values are obtained from calculations with 
  a large number of grid points.
  (b) Mulliken populations in the carbon chain.
  The sequential numbers 1 and 12 correspond to the most left
  and right hand side atoms in the central region $C$, respectively.}
\end{figure}

Since the majority part of the density matrix given by Eq.~(\ref{eq:c1-40}) is 
the equilibrium contribution, let us first discuss convergency of 
the equilibrium density matrix as a function of poles. 
The absolute error in $(E_{\rm tot}-E_{\rm self})$ of a carbon linear 
chain is shown in Fig.~2(a) as a function of the number of poles in order to 
illustrate the convergence property for the equilibrium density 
matrix under zero bias voltage,
where $(E_{\rm tot}-E_{\rm self})$ can be regarded as 
a conventional expression of the total energy in DFT, and 
the definitions of the two energy terms $E_{\rm tot}$ and 
$E_{\rm self}$ with the Fermi-Dirac function are given in Appendix A. 
For comparison the result calculated by a closed contour method is 
also shown.\cite{Brandbyge,ClosedContour}
One can see that the accuracy of $10^{-8}$ eV per atom is obtained 
using 140, 100, and 70 poles for the electronic temperature of 
300, 600, and 1200~K, respectively, while about 400 energy points 
are needed to obtain the same accuracy using the closed contour 
method at 600~K.\cite{Brandbyge} 
For most cases, we find that the convergence rate is similar to 
the case shown in Fig.~2(a). In general, 
the number of poles to achieve the accuracy of $10^{-8}$ eV per atom
must be proportional to the inverse of $T$, since the interval between 
neighboring poles of the continued fraction given by Eq.~(\ref{eq:c1-25}) is 
scaled by $k_{\rm B}T$. This fact implies that the computational effort 
increases as the electronic temperature decreases. However, we generally 
use electronic temperature from 300 to 1000~K for practical calculations, 
which means that the use of 100 poles is enough for practical purposes.
Therefore, it can be concluded that the most contribution of the density 
matrix can be very accurately evaluated with a small number of poles, 
i.e., 100.

To demonstrate the proper treatment of the boundary between 
the lead and the central regions in our implementation, 
in Fig.~2(b) we show a comparison between the conventional 
band structure and the EGF calculations with respect to DOS
of the carbon linear chain. The comparison provides a severe 
test to check whether the EGF method is properly implemented or not. 
It can be confirmed that DOS calculated by the EGF method is nearly 
equivalent to that by the conventional band structure calculation, 
which clearly shows the proper treatment of the boundary between 
the lead and the central regions in our scheme. 

As explained before, the integration of Eq.~(\ref{eq:c1-42}) 
required for the evaluation of the nonequilibrium term in the density matrix
has to be performed on the real axis with a small imaginary part 
because contour integration schemes may not be applied due to 
the non-analytic nature of the integrand. 
The treatment might suffer from numerical instabilities in the SCF 
iteration, since the integrand can rapidly vary due to the existence 
of poles of Green function located on the real axis. 
A remedy to avoid the numerical problem is to smear the Green 
function by introducing a relatively large imaginary part.\cite{Taylor,Brandbyge,RLi,Rocha} 

To investigate convergency of the nonequilibrium term in the density matrix, 
in Fig.~3(a) we show the absolute error in 
$(E_{\rm tot}-E_{\rm self})$ of the same infinite carbon chain 
as in Fig.~2 but under a finite bias voltage of 0.5 eV as 
a function of the number of regular grid points used for the 
evaluation of the nonequilibrium term in the density matrix given 
by Eqs.~(\ref{eq:c1-41}) and (\ref{eq:c1-42}).
We also tested the Gauss-Legendre quadrature for the 
integration of the nonequilibrium term besides the integration 
using the regular grid, but found that the convergence rate of 
the Gauss-Legendre quadrature is rather slower than the simple scheme
possibly due to the spiky structure of the integrand. 
Thus, we have decided to use the simple scheme using the regular grid. 
As expected, it turns out that the number of grid points to achieve 
the accuracy of $10^{-8}$ eV per atom increases as the imaginary part
becomes smaller. 
However, the accuracy of $10^{-8}$ eV is attainable using about 100 grid 
points in case of the imaginary part of 0.01 eV, while a few thousands 
grid points have to be used to achieve the same accuracy
for the imaginary part of 0.0001 eV. 

Although the accuracy of $10^{-8}$ eV can be achieved by introducing
the smearing scheme, however, one may consider that results can be 
affected by the introduction of an imaginary part.
In order to find a compromise between the accuracy and efficiency, 
the Mulliken population of the carbon linear chain under the finite 
bias voltage of 0.5 eV is shown in Fig.~3(b). 
We see that the use of the imaginary part 
of 0.01 eV gives a result comparable to that obtained by the use of 
0.0001 eV. Thus, the imaginary part of 0.01 eV can be a compromise 
between the accuracy and efficiency in this case. 
As a result, one can find that the energy points of 200 (100 and 100 
for the equilibrium and nonequilibrium
density matrices, respectively) are enough to achieve the accuracy 
of $10^{-8}$ eV per atom in $(E_{\rm tot}-E_{\rm self})$ at 600 K. 
However, it should be mentioned that a proper choice of the imaginary 
part must depend on the electronic structure of systems, and 
the careful consideration must be taken into account especially for 
a case that the spiky DOS appears in between two chemical potentials
of the leads, while for the equilibrium part of the density matrix 
the convergence property is insensitive to the electronic structure. 

\begin{figure}[t]
    \centering
    \includegraphics[width=8.0cm]{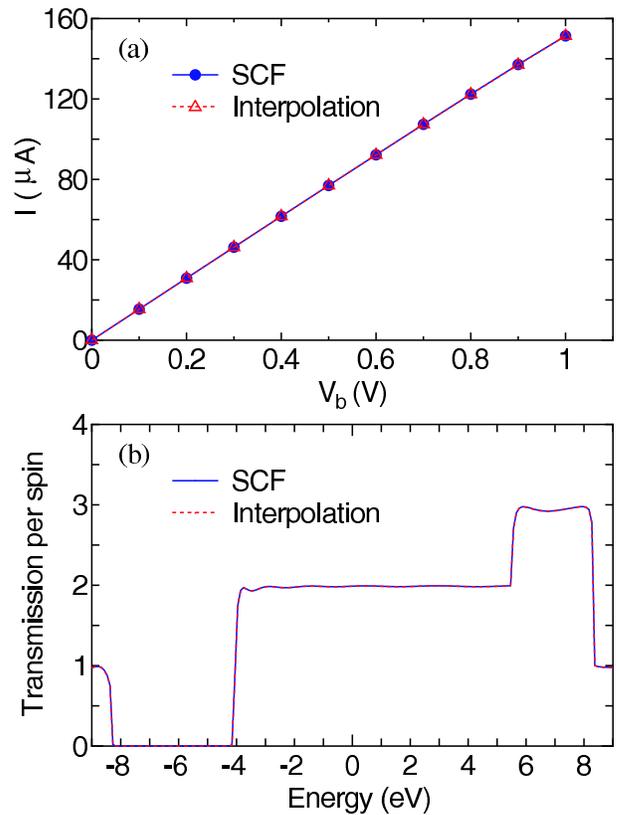}
    \caption{(Color online)
             (a) Currents of the linear carbon chain
             calculated by the SCF calculations (solid line)
             and the interpolation scheme (dotted line).
             (b) Transmission of the linear carbon chain under 
             a bias voltage of 0.3 V, 
             calculated by the SCF calculations (solid line)
             and the interpolation scheme (dotted line). 
             The imaginary part of 0.01 and the grid spacing of 0.01 eV
             are used for the integration of the nonequilibrium term 
             in the density matrix. 
             The other calculation conditions are same as for Fig.2(b)}
\end{figure}

We also note that the accurate evaluation of the density matrix 
makes the SCF calculation stable even under a finite bias voltage. 
In fact, for the case with the bias voltage of 0.5 V 
the number of the SCF iterations to achieve the residual norm 
of $10^{-11}$ for the charge density difference is 29 which 
is nearly equivalent to that, 30, for the zero bias case.

\subsection{Interpolation of the effect by the bias voltage}

Since for large-scale systems it is very time-consuming to perform 
the SCF calculation at each bias voltage, here we propose 
an interpolation scheme to reduce the computational cost in the 
calculations by the NEGF method. 
The interpolation scheme is performed in the following way: 
(i) the SCF calculations are performed 
for a few bias voltages which are selected in the regime of the bias 
voltage of interest.
(ii) when the transmission and current are calculated, 
a linear interpolation is made for the Hamiltonian block elements, 
$H_{\sigma,C}^{(\bf k)}$ and $H_{\sigma,R}^{(\bf k)}$, of 
the central scattering region and the right lead, and the chemical 
potential, $\mu_{R}$, of the right lead by 
\begin{eqnarray}
  H_{\sigma,C}^{(\bf k)} 
  &=& 
  \lambda H_{\sigma,C}^{({\bf k},1)} 
  + (1-\lambda) H_{\sigma,C}^{({\bf k},2)},
 \label{eq:c1-60}\\
 H_{\sigma,R}^{(\bf k)} 
  &=&
  \lambda H_{\sigma,R}^{({\bf k},1)} 
  + (1-\lambda) H_{\sigma,R}^{({\bf k},2)},
 \label{eq:c1-61}\\
 \mu_{R} &=& \lambda \mu_{R}^{(1)} + (1-\lambda) \mu_{R}^{(2)}, 
 \label{eq:c1-62}
\end{eqnarray}
where the indices $1$ and $2$ in the superscript mean that
the quantities are calculated or used at the corresponding bias voltages 
where the SCF calculations are performed beforehand. Note that 
it is also possible to perform the interpolation for ${\bf k}$-independent
Hamiltonian matrix elements instead of
Eqs.~(\ref{eq:c1-60}) and (\ref{eq:c1-61}).
In general, $\lambda$ should range from 0 to 1 for the moderate 
interpolation.  
A comparison between the fully self consistent and the interpolated 
results is shown with respect to the current and transmission 
in the linear carbon chain in Figs.~4(a) and (b). 
In this case, the SCF calculations at three bias voltages of 0, 0.5, 
and 1.0 V are performed, and the results at the other bias voltages 
are obtained by the interpolation scheme. For comparison we also 
calculate the currents via the SCF calculations at all the bias voltages. 
It is confirmed that the simple interpolation scheme gives notably
accurate results for both the calculations of the current and transmission. 
Although the proper selection of bias voltages used for the SCF calculations
may depend on systems, the result suggests that the simple scheme is 
very useful to interpolate the effect of the bias voltage while keeping 
the accuracy of the calculations.

\subsection{Applications}

\subsubsection{Zigzag graphene nanoribbons}

\begin{figure}[t]
    \centering
    \includegraphics[width=8.0cm]{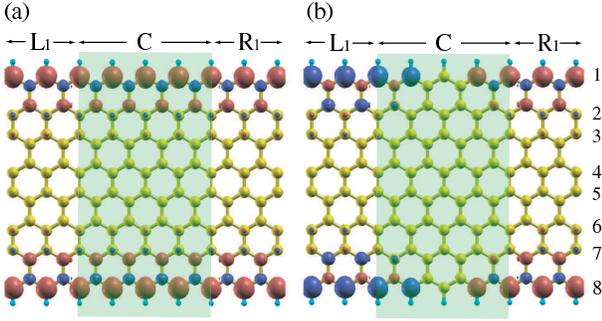}
    \caption{(Color online)
             8-ZGNR with (a) a FM junction and (b) an AFM junction
             together with the spatial distribution of the spin density
             at the source-drain bias voltage $V_{\rm b} = 0$ V. 
             The zigzag edges are terminated 
             by hydrogen atoms. The isosurface value of $\vert 0.002 \vert$ 
             is used for drawing the spin density.}
\end{figure}

As an illustration of our implementation, we investigate transport 
properties of zigzag graphene nanoribbons (ZGNRs) with different magnetic 
configurations. A characteristic feature in the band structure of ZGNR 
is the appearance of flat bands around X-point near the Fermi level, 
resulting in spin-polarization of associated states located at the 
zigzag edges.\cite{Fujita,Okada} 
Thus, so far several intriguing transport properties have been 
theoretically predicted 
especially for ZGNRs among GNRs by focusing on the spin polarized edge 
states.\cite{Son,Li,Kim,Abanin,Yazyev,Karpan1,Karpan2,Ezawa,Martins,Guo} 
For instance, it is found that ZGNRs might exhibit an extraordinary 
large magnetoresistance (MR) effect and a spin polarized current.\cite{Kim} 

\begin{figure}[b]
    \centering
    \includegraphics[width=8.0cm]{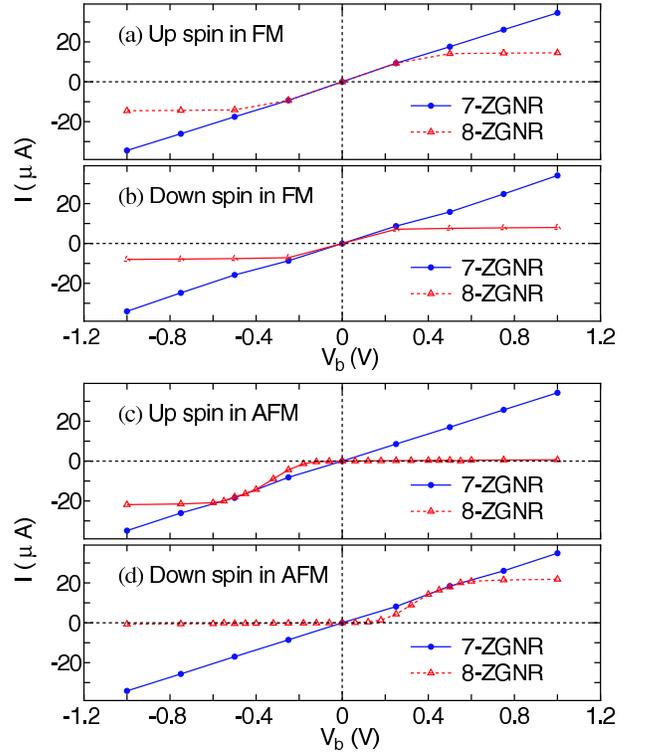}
    \caption{(Color online)
             $I$-$V_{\rm b}$ curves for (a) the up-spin and 
             (b) the down-spin states in the FM junction, and 
             (c) the up-spin and (d) the down-spin states 
             in the AFM junction of 7- and 8-ZGNRs.}
\end{figure}

Here we focus on the current-bias voltage ($I$-$V_{\rm b}$) 
characteristic of $7$- and $8$-ZGNRs with two magnetic configurations: 
ferromagnetic (FM) and antiferromagnetic (AFM) junctions 
as shown in Figs.~5(a) and (b), respectively, 
where the number, $7$ or $8$, is the number of carbon atom in the 
sublattice being across ZGNR along the lateral direction. 
The odd and even cases will also be referred to as {\it asymmetric} and
{\it symmetric}, respectively.
The extended central region $C$ consists of one sublattice and four unit 
cells, and contains 72 and 82 atoms for 7- and 8-ZGNRs, respectively. 
The poles of 100 is used for the evaluation of the equilibrium 
density matrix with the electronic temperature of 300 K, 
while the nonequilibrium term in the density matrix is evaluated 
using the simple quadrature method with the imaginary part of 0.01 eV 
and the grid spacing of 0.02 eV.   
The geometric structures used are optimized under the periodic 
boundary condition until the maximum force is less 
than $10^{-4}$ hartree/bohr.  
At each bias voltage the electronic structure of ZGNR is self-consistently
determined. 

\begin{figure}[t]
    \centering
    \includegraphics[width=8.0cm]{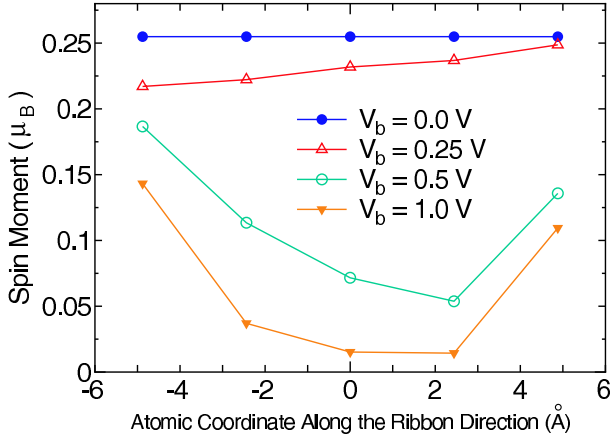}
    \caption{(Color online)
             Spin moments of the edge carbons in the central region $C$ 
             for the FM junction of 8-ZGNR under various
             applied source-drain bias voltages $V_{\rm b}$.}
\end{figure}

Figures~6(a) and (b) show the current-voltage ($I$-$V_{\rm b}$) curves 
for the up- and down-spin states in the FM junctions of 7- and 8-ZGNRs. 
It is found that the current for 7-ZGNR linearly depends on the bias 
voltage, while the current for 8-ZGNR is saturated 
at the bias voltage of about $\vert 0.5 \vert$. The distinct behavior
of 8-ZGNR from 7-ZGNR can be more definitely seen in the AFM junction as shown 
in Figs.~6(c) and (d). Interestingly, 8-ZGNR with the AFM junction
exhibits a diode behavior for the spin resolved current. 
Only the up-spin state contributes substantially to the current 
in the negative bias regime. 
In contrary in the positive bias regime only the down-spin 
state contributes to the current. It is worth mentioning that the effect 
can also be regarded as a spin filter effect. 
On the other hand, the $I$-$V_{\rm b}$ characteristics for 7-ZGNR is 
nearly equivalent to that for the FM junction. 
The considerable difference between 7- and 8-ZGNRs in the $I$-$V_{\rm b}$ 
characteristics can be attributed to the symmetry of two wave functions,
$\pi$ and $\pi^*$ states, around the Fermi level.
For 8-ZGNR the wave functions of the $\pi$ and $\pi^*$ states are 
antisymmetric and symmetric with respect to the $\sigma$ mirror plane 
which is the mid plane between two edges, respectively, while 
those wave functions for 7-ZGNR are neither symmetric nor antisymmetric.
From a detailed analysis\cite{Ozaki-ZGNR}, 
it can be concluded that the unique distinction in the $I$-$V_{\rm b}$ 
characteristics arises from an interplay between the symmetry of wave 
functions and band structures of ZGNRs. 
In addition, we find that spin moments are reduced by applying 
the finite bias voltage as shown in Fig.~7. 
Since the flat bands around X-point of the minority spin state are 
located about 0.25 eV above the Fermi level, the spin moments at 
the zigzag edges are largely reduced by increasing the occupation 
of the flat bands for the minority spin states when the bias voltage 
exceeds the threshold as shown in Fig.~7.\cite{Areshkin,Gunlycke} 
The details of the analysis on the unique spin diode and filter effect 
of ZGNRs is discussed elsewhere.\cite{Ozaki-ZGNR} 

\begin{figure}[b]
    \centering
    \includegraphics[width=8.5cm]{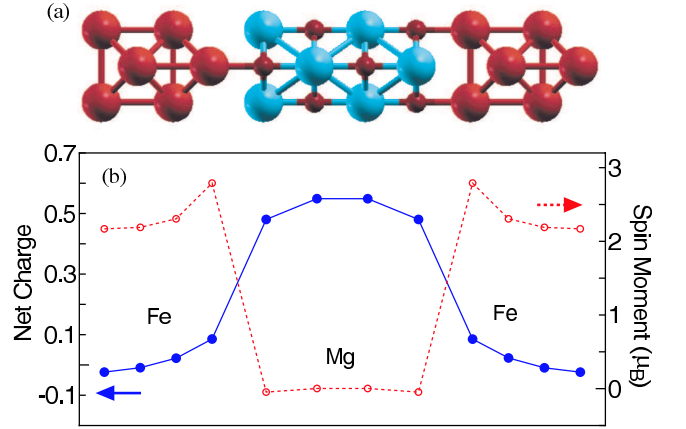}
    \caption{(Color online) (a) Calculation model for a tunneling junction 
             consisting of four MgO(100) layers, being the $C_0$ region,
             sandwiched by Fe(100), being the $L_0$ and $R_0$ regions.
             The red large and small and blue circles denote Fe, O, and 
             Mg atoms, respectively.
             (b) The net charge and spin magnetic moment of a Fe or Mg atom
             belonging to each layer in the parallel magnetic configuration 
             between the left and right leads. The position in the horizontal 
             axis exactly corresponds to that of the layer in the above figure.}
\end{figure}

\subsubsection{Fe/MgO/Fe tunneling junction}

The applicability of our implementation to bulk systems is demonstrated 
by an application to a tunneling junction consisting of MgO(100) layers 
sandwiched by iron. The magneto-tunneling junction was theoretically predicted
to exhibit a large tunneling-magnetoresistance (TMR),\cite{Butler}
and subsequently the TMR effect has been experimentally confirmed.\cite{Yuasa}   
We consider four MgO(100) layers sandwiched by iron with the (100) surface 
of which atomic configuration is shown in Fig.~8(a), where the lattice constant of 
the {\bf bc}-plane used is 2.866~\AA, and they are 2.866 and 4.054~\AA~in iron and 
MgO regions along the {\bf a}-axis, and the distance between the MgO and iron layers
is assumed to be 2.160~\AA.
The four MgO(100) layers corresponds
to the region $C_0$ in Fig.~1(a), and four Fe layers of the left and right hand
correspond the region $L_0$ and $R_0$, respectively. The SCF calculations were performed 
at 1000 K under zero bias voltage using {\bf k}-points of $7\times7$ and 
130 poles for the integration of the equilibrium density matrix.
It is found that obtaining the SCF is much harder than 
the case of ZGNR discussed before, and that a careful and modest treatment for the 
charge mixing is required.

\begin{figure}[t]
    \centering
    \includegraphics[width=8.0cm]{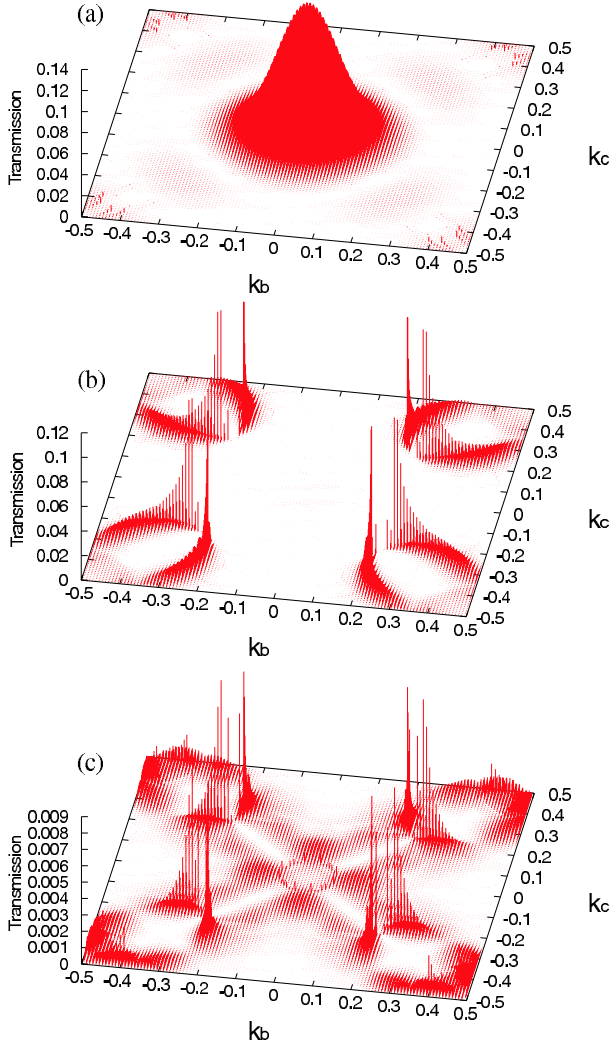}
    \caption{(Color online)
             {\bf k}-resolved Transmission at the chemical potential for 
             (a) the majority spin state of the parallel configuration, 
             (b) the minority spin state of the parallel configuration, and 
             (c) a spin state of the antiparallel configuration, respectively.
             For the calculations {\bf k}-points of $120\times 120$ were used.}
\end{figure}

As shown in Fig.~8(b), the net charge of iron atoms in the interfacial layer is positive
due to the coordination to an oxygen atom, and the reduction of electrons leads to 
the increase of the magnetic moment of iron atoms at the interfacial layer. 
The increase of the magnetic moment at the interfacial layer makes the distinction 
of the majority and minority spin states at the Fermi energy clear, i.e., the nature
of the majority and minority spin states at the Fermi energy can be assigned to $s$-
and $d$-states, respectively. The {\bf k}-resolved transmissions at the chemical potential
for the majority and minority spin states for the parallel magnetic configuration 
between the left and right leads are shown in Figs.~9(a) and (b), respectively. 
The large peak at the $\Gamma$ point in the majority spin state can be attributed to 
the $s$-state, while sharp peaks around four pillars come from the $d$-state as 
discussed in Ref.\cite{Butler}.
Note that the position of the sharp peaks is rotated by 45 degrees because of 
the unit cell rotated by 45 degrees compared to that in Ref.\cite{Butler}.
The conductances, $G_{\rm maj}^{\rm (p)}$ and $G_{\rm min}^{\rm (p)}$, for the majority and 
minority spin states, calculated from the average transmission integrated 
over the first Brillouin zone, are 11.99 and 2.82 $(\Omega^{-1} \mu {\rm m}^{-2})$, 
respectively, which implies that 
the tunneling junction may behave as a spin filter. The distinction in the conductance
should be attributed to decay properties of states in the insulating MgO region 
coupled with the two states.\cite{Butler} For the antiparallel magnetic configuration the 
{\bf k}-resolved transmission at the chemical potential is understood as 
a multiplication of the transmissions for the majority and minority spin states 
in the parallel configuration as shown in Fig.~9(c). The conductance, 
$G^{\rm (ap)}$, of the antiparallel magnetic 
configuration is 0.34 ($\Omega^{-1} \mu {\rm m}^{-2}$) which is 
smaller than those of the parallel case. By defining 
TMR$=(G_{\rm maj}^{\rm (p)}+G_{\rm min}^{\rm (p)} - 2 G^{\rm (ap)})/(2 G^{\rm (ap)})$,
we obtain TMR of 2082~\%, which is compared to 3700~\% for a 5 layer MgO case 
reported in Ref.\cite{Waldron}.

\subsubsection{LaMnO$_3/$SrMnO$_3$ superlattice}

When the transmission of a system with the periodicity along the {\bf a}-axis
as well as the periodicity of the {\bf bc}-plane is evaluated under zero bias voltage, 
it can be easily obtained by making use of the Hamiltonian calculated 
by the conventional band structure calculation without employing the Green 
function method described in the paper. 
This scheme enables us to explore transport properties for a wide variety 
of possible geometric and magnetic structures with a low computational cost, and
thereby can be very useful for many materials such as supperlattice structures. 
Once the Hamiltonian and overlap matrices are obtained from the conventional 
band structure calculation for the periodic structure, the transmission 
is evaluated by Eq.~(\ref{eq:c1-57}), where all the necessary information 
to evaluate Eq.~(\ref{eq:c1-57}) can be reconstructed by the result of the 
band structure calculation. As an example of the scheme, we calculate 
the conductance of a (LaMnO$_3$)/(SrMnO$_3$) superlattice with four different
magnetic structures, i.e., ferromagnetic, A-type, G-type, and C-type
antiferromagnetic configurations of Mn sites.\cite{Bhattacharya,Nakano} 

\begin{figure}[t]
    \centering
    \includegraphics[width=8.0cm]{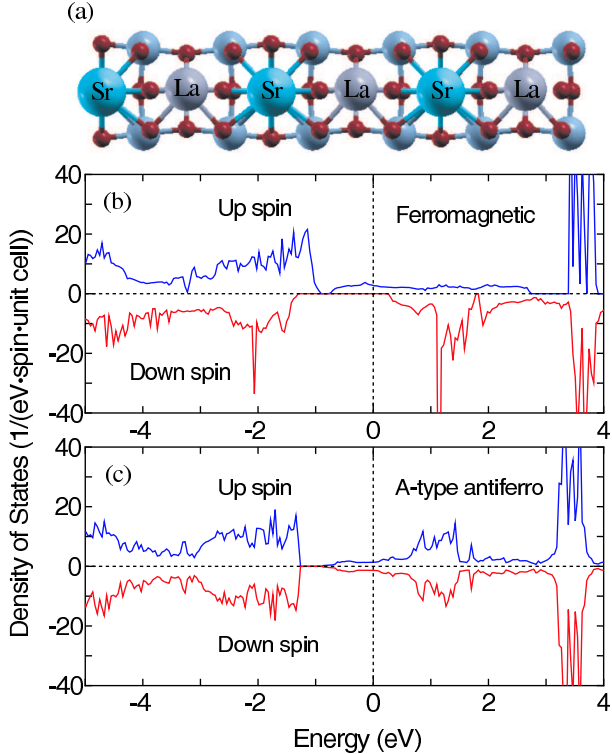}
    \caption{(Color online) (a) Optimized geometric structure of 
(LaMnO$_3$)/(SrMnO$_3$) superlattice with the ferromagnetic spin configuration of 
Mn sites. The middle sized blue and small red circles denote Mn and O atoms, 
while Sr and La atoms are denoted by the mark. Density of states (DOS) of
the (LaMnO$_3$)/(SrMnO$_3$) superlattice for 
(b) ferromagnetic and (c) A-type antiferromagnetic configurations.
  } 
\end{figure}

In recent years, it has been found experimentally that the superlattice structures 
consisting of LaMnO$_3$ and SrMnO$_3$ layers
exhibit a metal-insulator transition in terms of the layer thickness.\cite{Bhattacharya} 
Bhattacharya et al. fabricated (LaMnO$_3$)$_{2n}$/(SrMnO$_3$)$_{n}$ superlattices 
on a SrTiO$_3$ (001) substrate, and measured the in-plane resistivity as a function 
of temperature.\cite{Bhattacharya} 
The resistivity measurement indicates that the superlattices are 
metallic and insulating for $n \leq 2$ and $n \geq 3$, respectively.
On the other hand, the (LaMnO$_3$)$_{2}$/(SrMnO$_3$)$_{2}$ superlattices
fabricated on the same substrate by Nakano et al. exhibit a sample 
dependence in the resistivity measurement, i.e., 
one of three samples is metallic and the others are insulating.\cite{Nakano} 
They argued that the metallic behavior observed in the one sample may be attributed 
to a certain structural incompleteness in the superlattice structure, and 
that the ideal superlattice should become insulating based on their experimental 
results. A theoretical model calculation for the (LaMnO$_3$)$_{2n}$/(SrMnO$_3$)$_{n}$ 
superlattices suggests that the metal-insulator transition at $n=3$ can be explained 
by existence of the G-type antiferromagnetic barrier in the SrMnO$_3$ layers
sandwiched by the LaMnO$_3$ layers with the ferromagnetic configuration.\cite{Dong}
Since the analysis by Nakano et al. suggests that the charge transfer between 
the LaMnO$_3$ and SrMnO$_3$ layers is rather localized in the vicinity of 
the interface,\cite{Nakano} the model should be applicable to the case 
of (LaMnO$_3$)$_{2}$/(SrMnO$_3$)$_{2}$ without significantly depending on the ratio 
between the thicknesses of (LaMnO$_3$) and (SrMnO$_3$) layers, indicating
that (LaMnO$_3$)$_{2}$/(SrMnO$_3$)$_{2}$ is metallic. However, the naive consideration 
evidently contradicts the experimental result.\cite{Nakano}

As a first step towards comprehensive understanding of transport properties 
of the superlattice structures by the first principle calculations, we consider 
the simplest superlattice, i.e., (LaMnO$_3)/($SrMnO$_3$). In the calculations, 
the in-plane lattice constant is fixed to be 3.905 \AA~which is equivalent 
to that of the SrTiO$_3$ substrate. 
The out-of-plane lattice constant is assumed to be 7.735 \AA, since those are 
experimentally determined to be 3.959 \AA~and 3.776 \AA~for the LaMnO$_3$ 
and the SrMnO$_3$ layers, respectively, grown on the SrTiO$_3$ substrate 
and the average out-of-plane lattice constant for the superlattices is nearly 
equivalent to the average of the two values.\cite{Nakano} 
With those lattice constants internal structural parameters are optimized 
for each magnetic configuration without any constraint until the maximum force 
is less than $2.0\times 10^{-3}$ hartree/bohr. The optimized structure for
the ferromagnetic configuration is shown in Fig.~10(a). It is found that 
the position of oxygen atoms is largely distorted due to the different
ionic radii between La and Sr atoms, showing that Mn atoms are located 
in the center of each distorted octahedron. Also, bond angles of Mn-O-Mn 
are found to be 167.4 and 161.6 (degree) for the in-plane and out-of-plane, 
respectively. The total energies relative to the ferromagnetic configuration
are listed in Table~\ref{L1M1-TB}. The calculated ground state is the 
ferromagnetic configuration, and the A-type antiferromagnetic configuration 
lies just above 5 meV per formula unit. It may be considered that the nearly 
degeneracy between the two configurations corresponds to the neighborhood of 
the boundary at $x=0.5$ and $c/a=1$ in the phase diagram for the tetragonal 
La$_{1-x}$Sr$_x$MnO$_3$.\cite{Fang} 
The two configurations have both a metallic DOS, 
while the ferromagnetic configuration is half-metallic as shown in Figs.~10(b)
and (c), reflecting the large in-plane and out-of-plane conductances  
as shown in Table~\ref{L1M1-TB}. From a detailed analysis (not shown) of DOSs,  
we see that the electronic states at the Fermi level are composed of $e_g$ 
orbitals of Mn atoms and $p$ orbitals of oxygen atoms. Also, the Mulliken 
population analysis (not shown) suggests that the charge state of Mn atoms 
in the superlattice is in between those in the LaMnO$_3$ and SrMnO$_3$ bulks. 
This implies that the metallic band structures are induced by partial filling 
of the $e_g$ bands.
Since the bond angle of Mn-O-Mn for the out-of-plane is slightly acute, 
therefore, this may be attributed to reduction of the conductance of the 
ferromagnetic state in the out-of-plane as shown in Table~\ref{L1M1-TB}. 
A systematic study for the thicker cases and the effect of Coulomb 
interaction\cite{Han} in the $e_g$ orbitals is highly desirable, and the 
details will be discussed elsewhere.

\begin{table}[t]
\begin{center}
\caption{Total energy (meV) per formula unit, LaMnO$_3/$SrMnO$_3$,
and conductance $G$ ($\Omega^{-1} \mu {\rm m}^{-2}$) of the (LaMnO$_3)/($SrMnO$_3$) 
supperlattice with four different magnetic configurations, i.e., ferromagnetic (F), 
A-type (A), G-type (G), and C-type (C) antiferromagnetic configurations of Mn sites.
The total energy is measured relative to that of the ferromagnetic configuration.
$G_{\uparrow,{\rm in}}$ is the in-plane conductance for the up spin state, and 
the others are construed in the similar way. 
For the conductance calculations {\bf k}-points of $60\times 60$ were used.} 
\label{L1M1-TB}                     
\begin{tabular}{ccccc}
\hline
   & F & A & C & G  \\
\hline
 Energy                     & 0    & 5.0  & 163.8 & 248.2  \\
 $G_{\uparrow,{\rm in}}$    & 2262 & 1433 & 1169 & 1646  \\
 $G_{\downarrow,{\rm in}}$  & $1.82\times 10^{-2}$ & 1425  & 1105 & 1646 \\
 $G_{\uparrow,{\rm out}}$   & 1741 & 664 & 1127 & 678 \\
 $G_{\downarrow,{\rm out}}$ & $6.43\times 10^{-3}$ & 655 & 1128 & 677 \\
\hline
\end{tabular}
\end{center}
\end{table}

\subsection{Parallelization}

\begin{figure}[t]
    \centering
    \includegraphics[width=8.0cm]{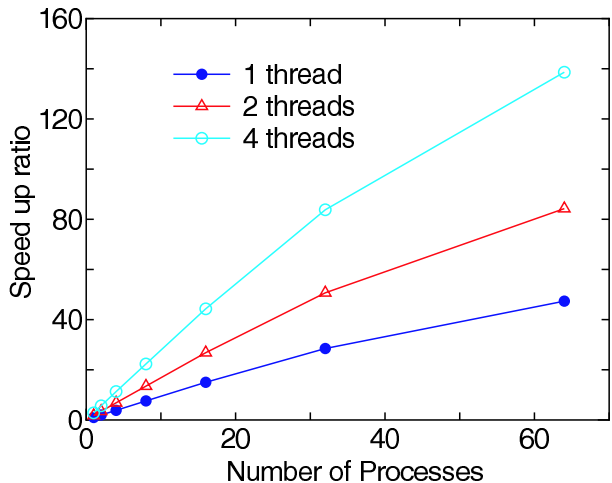}
    \caption{(Color online)
             Speed-up ratio in the parallel computation of
             the calculation of the density matrix
             for the FM junction of 8-ZGNR by a hybrid scheme 
             using MPI and OpenMP. The speed-up ratio is defined 
             by $T_1/T_{p}$, where $T_1$ and $T_p$ are the elapsed times
             by a single core and a parallel calculations.
             The cores used in the MPI and OpenMP parallelizations 
             are called {\it process} and {\it thread}, respectively.
             The parallel calculations were performed on a Cray XT5 machine
             consisting of AMD opteron quad core processors (2.3GHz).
             In the benchmark calculations, the number of processes is taken to
             be equivalent to that of processors. Therefore, in the parallelization 
             using 1 or 2 threads, 3 or 2 cores are idle in a quad core processor.}
\end{figure}

The computation in the the NEGF method can be parallelized in many aspects
such as {\bf k}-points, energies in the complex plane at which the Green 
functions are evaluated, spin index, matrix multiplications, 
and calculation of the inverse of the matrix. Here we demonstrate 
a good scalability of the NEGF in the parallel computation by a hybrid 
scheme using the message passing interface (MPI) and OpenMP which are 
used for inter-nodes and intra-node parallelization, respectively.
The Green function defined by Eq.~(\ref{eq:c1-6}) is specified by 
the {\bf k}-point, energy $Z$, and spin index $\sigma$. 
Since the calculation of the Green function specified by each set of 
three indices can be independently performed without any communication 
among the nodes, we parallelize triple loops corresponding to the three 
indices using MPI. Each node only has to calculate the Green functions 
for an allocated domain of the set of indices, and partly sum up 
Eq.~(\ref{eq:c1-26}) or (\ref{eq:c1-42}) in a discretized form. 
After all the calculations finish, a global summation among the nodes 
is required to complete the calculation of Eq.~(\ref{eq:c1-26}) or 
(\ref{eq:c1-42}), which, in most cases, is a very small fraction of the 
computational time even including the MPI communication among the nodes. 
Thus, the reduction of scalability for the parallelization of the three 
indices is mainly due to imbalance in the allocation of the domain of the 
set of indices. The imbalance can happen in the case that the number of 
combination for the three indices and the number of processes in the MPI 
parallelization are relatively small and large, respectively.
In addition to the three indices, one may notice that the matrix 
multiplications and the calculation of the matrix inverse can be 
parallelized, which are situated in the inner loops of the three indices. 
The evaluation of the central Green function given by Eq.~(\ref{eq:c1-6})
and the surface Green functions given by Eq.~(\ref{eq:c1-21}), 
and the self energies given by Eqs.~(\ref{eq:c1-7}) and (\ref{eq:c1-8}) 
are mainly performed by the matrix multiplications and the calculation 
of the matrix inverse. We parallelize these two computations using OpenMP
in one node. Since the memory is shared by threads in the node, 
the communication of the data is not required unlike the MPI parallelization. 
However, the conflict in the data access to the memory can reduce 
the scalability in the OpenMP parallelization. 
As a whole, in our implementation the {\bf k}-point, energy $Z$, and 
spin index $\sigma$ are parallelized by MPI, and the matrix multiplications 
and the calculation of the matrix inverse are parallelized by OpenMP.

In Fig.~11 we show the speed-up ratio in the elapsed time for the evaluation 
of the density matrix of 8-ZGNR under 
a finite bias voltage of 0.5 eV. The geometric and magnetic structures and 
computational conditions for 8-ZGNR are same as before. The energy points of 197 
(101 and 96 for the equilibrium and nonequilibrium terms, 
respectively) are used for the evaluation of the density matrix. 
Only the $\Gamma$ point is employed for the {\bf k}-point sampling, and 
the spin polarized calculation is performed. Thus, the combination of 394
for the three indices are parallelized by MPI. It is found that the speed-up 
ratio of the flat MPI parallelization, corresponding to 1 thread, 
reasonably scales up to 64 processes. Furthermore, it can be seen that the 
hybrid parallelization, corresponding to 2 and 4 threads, largely improves 
the speed-up ratio. By fully using 64 quad core processors, corresponding to 
64 processes and 4 threads, the speed-up ratio is about 140, demonstrating 
the good scalability of the NEGF method.

\section{CONCLUSIONS}

We have presented an efficient and accurate implementation of 
the NEGF method for electronic transport calculations
in combination with DFT using pseudo-atomic orbitals and 
pseudopotentials. In the implementation, we have developed 
accurate methods for the evaluation of the density matrix and 
the treatment of the boundary between the scattering region and the leads. 

A contour integration method with a continued fraction representation 
of the Fermi-Dirac function has been successfully applied for 
the evaluation of the equilibrium term in the density matrix,
which evidently outperforms the previous method,\cite{Brandbyge} 
while a simple quadrature scheme on the real axis with a small imaginary part
is employed for that for the nonequilibrium term in the density matrix. 
It has been demonstrated by numerical calculations that 
the accuracy of $10^{-8}$ eV per atom in $(E_{\rm tot}-E_{\rm self})$ is 
attainable using the energy points of 200 in the complex plane even 
under a finite bias voltage of 0.5 V at 600 K. 
However, the evaluation of the nonequilibrium density matrix still 
requires a careful treatment where the pole structure of the Green 
functions has to be smeared by introducing a finite imaginary part.
The numerical calculations suggest that the number of energy points 
required for the convergence can be largely reduced by introducing 
the imaginary part of 0.01 eV without largely changing the calculated 
results in a practical sense.
We also note that the accurate evaluation of the density 
matrix provides another advantage that the SCF calculations 
even under a finite bias voltage smoothly converge in a similar fashion 
as the conventional band structure calculation does. 

We have also developed an efficient method for calculating the Hartree 
potential by a combination of the two dimensional FFT and a finite difference 
method without any ambiguity in reproducing the boundary conditions.
In addition, a careful evaluation of the charge density near the boundary 
between the scattering region and the leads is presented in order to 
avoid the spurious scattering accompanied by 
the inaccurate construction of the charge density. 
The proper treatment for the charge evaluation in our implementation 
can definitely be verified by a comparison between the conventional band 
structure calculation and the EGF method with respect to the DOS of the 
carbon chain.  

Finally, we have demonstrated the applicability of our implementation 
by calculations of spin resolved $I$-$V_{\rm b}$ characteristics of ZGNRs, 
showing that the $I$-$V_{\rm b}$ characteristics depend on the symmetry 
of ZGNR, and that the symmetric ZGNR exhibits a unique spin diode and filter 
effect. 
Also, the applicability of our implementation to bulk systems is demonstrated 
by applications to a Fe/MgO/Fe tunneling junction 
and a LaMnO$_3/$SrMnO$_3$ superlattice. 
Based on the above discussions and the good parallel efficiency in the hybrid 
parallelization shown in the study, it is concluded that our implementation 
of the NEGF method can be applicable to challenging problems related to
large-scale systems, and can be a starting point, 
apart from numerical spurious effects, to include many body effects
beyond the one particle picture in the electronic transport.

\acknowledgments

The authors would like to thank H. Kondo for providing a prototype
FORTRAN code of the NEGF method. 
This work is partly supported by CREST-JST, the Next 
Generation Supercomputing Project, Nanoscience Program,
and NEDO (as part of the Nanoelectronics project).

\appendix 

\section{Total energy functional for the equilibrium state}

Let us introduce the following density functional which may define
the total energy of the central region $C$ being a part of 
the extended system at equilibrium
with a common chemical potential $\mu$:
\begin{eqnarray}
  E_{\rm tot}
  = 
  E_{\rm kin}
  + 
  E_{\rm ext}
  + 
  E_{\rm ee}
  + 
  E_{\rm xc}
  + 
  E_{\rm self},
 \label{eq:a-1}
\end{eqnarray}
where $E_{\rm ext}$ is the Coulomb interaction energy between electrons
and the external potential of the central region $C$ given by 
\begin{eqnarray}
 E_{\rm ext}
 &=& 
  \int dr^3
   n({\bf r})
   v_{\rm ext}({\bf r})
  \label{eq:a-2},
\end{eqnarray}
and $E_{\rm ee}$ is the Hartree energy defined by 
\begin{eqnarray}
 \nonumber
 E_{\rm ee}
 &=& 
  \frac{1}{2}
  \int \int dr^3 dr'^3
   \frac{\left(n({\bf r})+n'({\bf r})\right)
         \left(n({\bf r}')+n'({\bf r}')\right)}
        {\vert {\bf r} - {\bf r}' \vert}
  \label{eq:a-3},\\
\end{eqnarray}
where $n'$ is an additional electron density which arises
from the boundary condition between the central and lead regions. 
In fact, the additional electron density $n'$ can be obtained by 
back Fourier transforming Eq.~(\ref{eq:c1-39}) divided by $4\pi(\Delta x)^2$
within our treatment. The last term of Eq.~(\ref{eq:a-1}), 
$E_{\rm self}$, is a self energy density functional, which 
may correspond to the energy contribution from the self energy
due to the semi infinite leads, given by 
\begin{eqnarray}
 E_{\rm self}
 &=& 
  {\rm Tr}\left[
   -\frac{2}{\pi}
   {\rm Im}
   \int_{-\infty}^{\mu}
   dE
    G_{C}(E^+)
    \Lambda(E)
   \right]
  \label{eq:a-4}
\end{eqnarray}
with
\begin{eqnarray}
  \Lambda(E)
    = 
    \Sigma(E^+)
      - E\frac{\partial \Sigma(E^+)}{\partial E},
  \label{eq:a-5}
\end{eqnarray}
where $E^+\equiv E+i0^+$, the factor of 2 is due to the spin multiplicity, 
and the self energy $\Sigma$ is the sum of the self energies arising from 
the left and right leads. 
Although we neglect the spin and {\bf k}-dependency on the formulation for the 
simplicity throughout the Appendix, its generalization with the dependency 
is straightforward. The kinetic energy $E_{\rm kin}$ can be evaluated 
by Eq.~(\ref{eq:c1-6}) as the band energy $E_{\rm band}$ minus double counting 
corrections as follows:
\begin{eqnarray}
 \nonumber
 E_{\rm kin}
 &=& 
  {\rm Tr}\left[
   -\frac{2}{\pi}
   {\rm Im}
   \int_{-\infty}^{\mu}
   dE
   G_{C}(E^+)
   H_{C,{\rm kin}}
  \right]\\
 \nonumber
 &=&
  E_{\rm band}
  -
  \int dr^3
   n({\bf r})
   v_{\rm eff}({\bf r})\\
 \nonumber
 &&
  - 
  {\rm Tr}\left[
   -\frac{2}{\pi}
   {\rm Im}
   \int_{-\infty}^{\mu}
   dE
   G_{C}(E^+)
   \Sigma(E^+)
  \right]
  \label{eq:a-6},\\
\end{eqnarray}
where $E_{\rm band}$ is defined by 
\begin{eqnarray}
  E_{\rm band}
  = 
  {\rm Tr}\left[
   -\frac{2}{\pi}
   {\rm Im}
   \int_{-\infty}^{\mu}
   dE
   E
   G_{C}(E^+)
   S_{C}
  \right].
  \label{eq:a-7}
\end{eqnarray}
It is noted that the last term in Eq.~(\ref{eq:a-6}) cancels
the contribution from the first term in Eq.~(\ref{eq:a-5}). 
The effective potential $v_{\rm eff}$ in the second term of Eq.~(\ref{eq:a-6})
will be defined later. 
Also, the exchange-correlation energy $E_{\rm xc}$ in Eq.~(\ref{eq:a-1}) 
is considered to be a density functional such as local density 
approximations (LDA) and generalized gradient approximations (GGA) evaluated
using electron density $n$ in the central region $C$. 

We now consider the variation of the total energy $E_{\rm tot}$ with respect 
to $n$.
The variations of $E_{\rm ext}$ and $E_{\rm ee}$ are simply given by 
\begin{eqnarray}
  \delta[E_{\rm ext}]
  = 
  \int dr^3
   \delta n({\bf r})
   v_{\rm ext}({\bf r})
  \label{eq:a-8}
\end{eqnarray}
and 
\begin{eqnarray}
  \delta[E_{\rm ee}]
  = 
   \int dr^3
   \delta n({\bf r})
   \int dr'^3
   \frac{n({\bf r}')+n'({\bf r}')}
        {\vert {\bf r} - {\bf r}' \vert}.
  \label{eq:a-9}
\end{eqnarray}
By noting the Dyson equation and 
$G_CS_CG_C=-\frac{\partial G_C}{\partial E}+G_C\frac{\partial \Sigma}{\partial E}G_C$
which are both derived from Eq.~(\ref{eq:c1-6}), and ${\rm Tr}(AB)={\rm Tr}(BA)$, 
the variation of $E_{\rm band}$ is given by two contributions:
\begin{eqnarray}
  \nonumber
 \lefteqn{
  \delta[E_{\rm band}]
 }\\
 \nonumber
 && = 
  \int dr^3
  \delta n({\bf r})
  {\rm Tr}\left[
   -\frac{2}{\pi}
   {\rm Im}
   \int_{-\infty}^{\mu}
   dE\right.\\
  \nonumber
  &&
  \qquad\qquad
  \left.
  \times
   E
   G_{C}(E^+)
   \frac{\delta H_{v}}
        {\delta n({\bf r})} 
   G_{C}(E^+)
   S_{C}
  \right]\\
 \nonumber
 && = 
  \int dr^3
  \delta n({\bf r})
  {\rm Tr}\left[
   -\frac{2}{\pi}
   {\rm Im}
   \int_{-\infty}^{\mu}
   dE
   \right.\\
  &&
 \nonumber
   \left.
   \qquad\qquad
   \times
   E\left(
   \frac{-\delta H_{v}}
        {\delta n({\bf r})} 
   \right)
   \frac{\partial G_{C}(E^+)}{\partial E}
  \right]\\
  \nonumber
  &&
   \quad
  +
  \int dr^3
  \delta n({\bf r})
  {\rm Tr}\left[
   -\frac{2}{\pi}
   {\rm Im}
   \int_{-\infty}^{\mu}
   dE
  \right.\\
  \nonumber
  && 
   \qquad
   \qquad
   \left.
   \times
   E
   G_{C}(E^+)
   \frac{\delta H_{v}}{\delta n({\bf r})}
   G_{C}(E^+)
   \frac{\partial \Sigma(E^+)}{\partial E}
  \right].\\
  \label{eq:a-10}
\end{eqnarray}
The trace in the first term of Eq.~(\ref{eq:a-10}) can be transformed 
by considering a partial integral and assuming the system to be insulating
as follows:
\begin{eqnarray}
  \nonumber
 \lefteqn{
  {\rm Tr}\left[
   -\frac{2}{\pi}
   {\rm Im}
   \int_{-\infty}^{\mu}
   dE
   E
   \left(
   \frac{-\delta H_{v}}{\delta n({\bf r})} 
   \right)
   \frac{\partial G_{C}(E^+)}{\partial E}
  \right]
  }\\
  \nonumber
  &&=
  -{\rm Tr}\left[
   -\frac{2}{\pi}
   {\rm Im}
   \int_{-\infty}^{\mu}
   dE
   \left(
   \frac{-\delta H_{v}}{\delta n({\bf r})} 
   \right)
   G_{C}(E^+) 
  \right]\\
  &&=
  \int dr'^3
  n({\bf r}')
  \frac{\delta v_{\rm eff}({\bf r}')}{\delta n({\bf r})}.
  \label{eq:a-11}
\end{eqnarray}
It is also noted that the second term in Eq.~(\ref{eq:a-10})
cancels the variation of the contribution from the second term in 
Eq.~(\ref{eq:a-5}). 
The variation of the second term in Eq.~(\ref{eq:a-6}) is easily 
found as 
\begin{eqnarray}
  \nonumber
  \delta
  \left[
  \int dr^3
   n({\bf r})
   v_{\rm eff}({\bf r})
  \right]
  &=&
  \int dr^3
  \delta n({\bf r})
  v_{\rm eff}({\bf r})\\
  \nonumber
  &+&
  \int dr^3
  \delta n({\bf r})
  \int dr'^3
  n({\bf r}')
  \frac{\delta v_{\rm eff}({\bf r}')}{\delta n({\bf r})}.\\
  \label{eq:a-12}
\end{eqnarray}
Thus, it turns out using Eqs.~(\ref{eq:a-8})-(\ref{eq:a-12}) that 
the variation of the total energy $E_{\rm tot}$ is given by 
\begin{eqnarray}
  \nonumber
  \lefteqn{
  \delta\left[E_{\rm tot}\right]
  }\\
  \nonumber
  &&
  =
   \int dr^3
   \delta n({\bf r})\\
  \nonumber
  &&
  \times
  \left(
  -v_{\rm eff}({\bf r})
  +v_{\rm ext}({\bf r})
  + 
   \int dr'^3
   \frac{n({\bf r}')+n'({\bf r}')}
        {\vert {\bf r} - {\bf r}' \vert}
  + \frac{\delta E_{\rm xc}}{\delta n({\bf r})}
  \right).\\
  \label{eq:a-13}
\end{eqnarray}
Letting $\delta\left[E_{\rm tot}\right]$ be zero
so that the variation of the total energy $E_{\rm tot}$ can be always zero
with respect to $n$, we obtain a form of the effective potential:
\begin{eqnarray}
  \nonumber
  v_{\rm eff}({\bf r})
  =
  v_{\rm ext}({\bf r})
  + 
   \int dr'^3
   \frac{n({\bf r}')+n'({\bf r}')}
        {\vert {\bf r} - {\bf r}' \vert}
  + \frac{\delta E_{\rm xc}}{\delta n({\bf r})}.\\
  \label{eq:a-14}
\end{eqnarray}
It is found that the effective potential takes the same form as 
in the Kohn-Sham method.\cite{Kohn} 
The fact implies that the self consistent solution of the Green
function under the zero bias condition may correspond to 
the minimization of the total energy functional defined 
by Eq.~(\ref{eq:a-1}), since in practice the 
Green function defined by Eq.~(\ref{eq:c1-6}) 
is calculated using the effective potential given by Eq.~(\ref{eq:a-14})
as a consequence of combining the NEGF method with DFT. 

The generalization of the functional to the metallic 
case with a finite temperature and the nonequilibrium state might 
be an important direction in the future study so that 
forces on atoms can be {\it variationally} calculated from 
the functional, 
since the existence of a variational functional has been recently 
suggested for the nonequilibrium steady state.\cite{Todorov,Ventra,Ventra2}

\section{Energy density matrix}

The equilibrium energy density matrix $e_{\sigma,{\bf R}_{\rm n}}^{\rm (eq)}$, 
where one of the associated basis orbitals is in the central cell 
and the other is in the cell denoted by ${\bf R}_{\rm n}$,
is calculated using the contour integration method applied to
the equilibrium density matrix as follow:
\begin{eqnarray}
   e_{\sigma,{\bf R}_{\rm n}}^{\rm (eq)} 
   = 
    \frac{1}{V_{\rm c}}
    \int_{\rm BZ} dk^3
    \left(
    e_{\sigma,+}^{(\bf k)} - e_{\sigma,-}^{(\bf k)}
    \right)
    {\rm e}^{-i{\bf k}\cdot {\bf R}_{\rm n}}
  \label{eq:c1-27}
\end{eqnarray}
with 
\begin{eqnarray}
    e_{\sigma,\pm}^{(\bf k)}
    = 
    \frac{i}{2\pi}
    \int_{-\infty}^{\infty}
    dE
    EG_{\sigma,C}^{(\bf k)}(E\pm i0^+)
    f(E-\mu).
    \label{eq:c1-28}
\end{eqnarray}
If the Hamiltonian and overlap matrices are ${\bf k}$-independent, 
as well as Eq.~(\ref{eq:c1-22}), Eq.~(\ref{eq:c1-28}) can be 
simplified to:
\begin{eqnarray}
   e_{\sigma,0} 
   = 
  {\rm Im}\left[
    -\frac{1}{\pi}
    \int_{-\infty}^{\infty}
    dE
    E
    G_{\sigma,C}(E+i0^+)
    f(E-\mu)
    \right].
  \label{eq:c1-29}
\end{eqnarray}
For the general case with the ${\bf k}$-dependent Hamiltonian and overlap 
matrices, Eq.~(\ref{eq:c1-28}) is evaluated by the contour integration 
method with the special form of Fermi-Dirac function given by 
Eq.~(\ref{eq:c1-25}) as follows:
\begin{eqnarray}
   \nonumber
   e_{\sigma,\pm}^{(\bf k)}
    =  
   \pm\frac{1}{4}\mu^{({\bf k},1)}_{\sigma}
   \pm
   \frac{1}{2}
   \gamma_0 
   \mu^{({\bf k},0)}_{\sigma}
   \mp 
    \frac{1}{\beta} \sum_{p=1}^{N_p} 
    G_{\sigma,C}^{(\bf k)}(\alpha_p)R_p \alpha_p
    \label{eq:c1-30}\\
\end{eqnarray}
with 
\begin{eqnarray}
  \gamma_0 
  =  
  \frac{2}{\beta}
  \sum_{p=1}^{N_p} R_p,
  \label{eq:c1-31}
\end{eqnarray}
where $\mu^{({\bf k},1)}_{\sigma}$ is the first order 
moment of the Green function $G_{\sigma,C}^{(\bf k)}$.
In Eq.~(\ref{eq:c1-30}), a term,  
$\frac{i}{2\pi}\lim_{R\to \infty} R\mu^{({\bf k},0)}_{\sigma}$,
which appears mutually for $e_{\sigma,\pm}^{(\bf k)}$, 
is omitted, since the diverging terms cancel each other out
in Eq.~(\ref{eq:c1-27}). 
By making use of the moment representation of the Green function,\cite{Ozaki-FD} 
the following simultaneous linear equation is derived 
for the zero and first order moments:
\begin{eqnarray}
 \left( 
   \begin{array}{cc}
      1  & z_0^{-1} \\
      1  & z_1^{-1} \\
   \end{array}
 \right)
 \left( 
   \begin{array}{c}
     \mu^{({\bf k},0)}_{\sigma}\\
     \mu^{({\bf k},1)}_{\sigma}\\
   \end{array}
 \right)
  = 
 \left( 
   \begin{array}{c}
     z_0 G_{\sigma,C}^{(\bf k)}(z_0)\\
     z_1 G_{\sigma,C}^{(\bf k)}(z_1)\\
   \end{array}
 \right),
  \label{eq:c1-32}
\end{eqnarray}
Letting $z_0$ and $z_1$ be $iR$ and $-R$, respectively, 
the zero and first order moments can be evaluated by 
 \begin{eqnarray}
  \mu^{({\bf k},0)}_{\sigma}
  &=&
   \frac{R}{1-i}
   \left(
     G_{\sigma,C}^{(\bf k)}(iR)
    -
     G_{\sigma,C}^{(\bf k)}(-R)
   \right),
  \label{eq:c1-33}\\
  \mu^{({\bf k},1)}_{\sigma}
  &=&
   \frac{iR^2}{1+i}
   \left(
     iG_{\sigma,C}^{(\bf k)}(iR)
     +
      G_{\sigma,C}^{(\bf k)}(-R)
   \right),
  \label{eq:c1-34}
\end{eqnarray}
where $R$ is a large real number so that the higher order moments
can be neglected.

For the nonequilibrium Green function, the nonequilibrium
contribution $\Delta e_{\sigma}$  
in the the energy density matrix $e_{\sigma}^{\rm (neq)}$ 
can be calculated using the simple quadrature
scheme in the same way as for the nonequilibrium term 
in the density matrix.


\begin{thebibliography}{99}
\bibitem{ClosedContour}
We used $\Delta=1.2$~eV, EB=27.21 eV, and $\gamma=40k_{\rm B}T$, 
being the parameters in the closed contour integration
depicted in Fig.~2 of Ref.\cite{Brandbyge}. 
The value for $\Delta$ gives four enclosed Matsubara poles for $T=600$~K.
The path $L$ toward the positive direction is extended 
by $40k_{\rm B}T$ from the chemical potential. 
The paths $L$ and $C$ are integrated using the Gauss-Legendre
quadrature scheme, respectively. It turns out that 
40 grid points for the path $C$ is enough to achieve 
the fully convergent result within double precision. 
So, we used 40 grid points for the path $C$, and only changed 
the number of grid points for the path $L$ in the calculation
for Fig.~2(a) of Ref.\cite{Brandbyge}. Therefore, the number of energy points
in the closed contour integration 
is given by 4 (the enclosed Matsubara poles) plus 40 (the path $C$)
plus the number of grid points for the path $L$.
\end{thebibliography}
\end{document}